\documentclass[12pt]{article}

\usepackage{newtxtext,newtxmath}

\usepackage{graphicx}
\usepackage{multirow}
\usepackage[letterpaper,margin=1in]{geometry}
\usepackage[round, numbers]{natbib}
\usepackage{rotating}

\renewenvironment{abstract}
	{\quotation}
	{\endquotation}

\date{}

\makeatletter
\renewcommand{\fnum@figure}{\textbf{Figure \thefigure}}
\renewcommand{\fnum@table}{\textbf{Table \thetable}}
\makeatother

\usepackage{scicite}

\usepackage{url}
\usepackage[normalem]{ulem}
\usepackage{orcidlink}

\newcommand{\subtitle}[1]{%
  \begin{center}
    \large\textbf{#1}
  \end{center}
}

\def\scititle{
	Discovery of a likely supermassive neutron star in a remarkably wide binary \\
    Short title: supermassive neutron star candidate
}
\title{\bfseries \boldmath \scititle}

\author{
    Xinlin Zhao$^{1\dagger}$~\orcidlink{0009-0005-5459-7433},
    Song Wang$^{2,3\ast}$~\orcidlink{0000-0003-3116-5038},
    Guang-Yao Xiao$^{4,5\dagger}$~\orcidlink{0000-0001-6753-4611},
    Meng Sun$^{2,6}$,\and
    Roberto Soria$^{7,8}$,
    Kyle Akira Rocha$^{6,9,10}$,
    Seth Goggage$^{6,10}$,
    Pengfei Wang$^{11}$,\and
    Hongliang Yan$^{2,3}$~\orcidlink{0000-0002-8609-3599},
    Xuan Mao$^{12,13,14}$,
    Yang Huang$^{13}$,
    Maosheng Xiang$^{2,3}$,\and
    Shuai Liu$^{2}$~\orcidlink{0000-0001-5193-1727},
    Dan Qiu$^{15,12}$,
    Meng Zhang$^{2}$,
    Xue Li$^{13}$~\orcidlink{0009-0007-4501-4376},
    Fabo Feng$^{4,5}$,\and
    Xihong Luo$^{1}$,
    Xiaohong Yang$^{1}$,
    Jifeng Liu$^{2,13,16\ast}$~\orcidlink{0000-0002-2874-2706}\and
    \small$^{1}$Department of Physics and Chongqing Key Laboratory for Strongly Coupled Physics, \\
    \small Chongqing University, Chongqing, 401331, China.\and
    \small$^{2}$National Astronomical Observatories, Chinese Academy of Sciences, Beijing, 100101, China.\and
    \small$^{3}$Institute for Frontiers in Astronomy and Astrophysics, Beijing Normal University, Beijing, 102206, China.\and
    \small$^{4}$Tsung-Dao Lee Institute, Shanghai Jiao Tong University, Shanghai, 201210, China.\and
    \small$^{5}$School of Physics and Astronomy, Shanghai Jiao University, Shanghai, 200240, China.\and
    \small$^{6}$Center for Interdisciplinary Exploration and Research in Astrophysics (CIERA), \\
    \small Northwestern University, Evanston, 60201, USA.\and
    \small$^{7}$INAF--Osservatorio Astrofisico di Torino, Strada Osservatorio 20, I-10025 Pino Torinese, Italy.\and
    \small$^{8}$Sydney Institute for Astronomy, School of Physics A28, The University of Sydney, Sydney, NSW 2006, Australia.\and
    \small$^{9}$Department of Physics \& Astronomy, Northwestern University, Evanston, 60208, USA.\and
    \small$^{10}$NSF-Simons AI Institute for the Sky (SkAI), 172 E. Chestnut St., Chicago, 60611, USA.\and
    \small$^{11}$State Key Laboratory of Radio Astronomy and Technology, National Astronomical Observatories, \\
    \small Chinese Academy of Sciences, Beijing, 100101, China.\and
    \small$^{12}$Key Laboratory of Space Astronomy and Technology, National Astronomical Observatories, \\
    \small Chinese Academy of Sciences, Beijing, 100101, China.\and
    \small$^{13}$School of Astronomy and Space Sciences, University of Chinese Academy of Sciences, Beijing 100049, China.\and
    \small$^{14}$INAF – Osservatorio Astronomico di Palermo, Piazza del Parlamento 1, 90134 Palermo, Italy \\
    \small$^{15}$Kavli Institute for Astronomy and Astrophysics, Peking University, Beijing, 100871, China\and
    \small$^{16}$New Cornerstone Science Laboratory, National Astronomical Observatories, \\
    \small Chinese Academy of Sciences, Beijing, 100101, China.\and
	\small$^\ast$Corresponding author. Email: songw@bao.ac.cn; jfliu@nao.cas.cn \\
    \small$^\dagger$These authors contributed equally to this work.
}

\begin{document}

\maketitle

\noindent\subtitle{Abstract}

\begin{abstract} \bfseries \boldmath
Supermassive neutron stars ($M \gtrsim 2\ M_{\odot}$) provide stringent constraints on the equation of state of dense matter.
We report the discovery of a wide binary system composed of a red giant star and an invisible companion with an orbital period $P\approx 1218\pm3$ days and a comparatively low eccentricity $e\approx 0.09\pm0.01$ (with $1 \sigma$ uncertainty).
By combining radial velocity measurements from the Large Sky Area Multi-Object Fiber Spectroscopic Telescope and the Apache Point Observatory Galactic Evolution Experiment (APOGEE) with astrometric data from {\it Gaia} DR2 and DR3 catalogues, we derive a Keplerian orbit with an inclination angle $i\approx73^{\circ}\pm11^{\circ}$.
The unseen companion has a mass in the range of 1.8--2.8 $M_{\odot}$ (spanning the full spread of all estimates), with a best-fit value of  $2.20^{+0.14}_{-0.13}$ $M_{\odot}$.
This potentially places the unseen companion within the mass gap between neutron stars and black holes.
Spectral disentangling using high-resolution near-infrared APOGEE spectra shows no absorption features from an additional component, suggesting the presence of a compact object: either a supermassive neutron star or a low-mass black hole.
The wide, low-eccentricity orbit is difficult to explain through standard binary evolution channels, pointing to the need for improved theoretical models to account for such systems.
\end{abstract}

\subsection*{Introduction}

The physical and orbital properties of neutron stars (NSs) in binaries encode their formation and evolutionary history.
For example, non-recycled NSs typically exhibit eccentric orbits and a mass distribution peaking at 1.33 $M_{\odot}$, whereas recycled NSs have circular orbits and a mass distribution peaking at 1.54 $M_{\odot}$ \citep{2016ARA&A..54..401O}.
A recent study further proposed a bimodal distribution in the recycled millisecond pulsar (MSP) population, with two peaks at 1.39 $M_{\odot}$ and 1.81 $M_{\odot}$ \citep{2016arXiv160501665A}. 
These findings point to a diversity of mass transfer histories during their evolution.
Additionally, to date, only a handful of NSs with masses exceeding 2 $M_{\odot}$ have been reported \citep{2003ApJ...596L..63A,2008ApJ...675..670F,2011ApJ...730...25R,2015A&A...577A.130F,2021ApJ...915L..12F,2024Sci...383..275B}, which can provide essential limits on the maximum NS mass and thus place strong constraints on the NS equation of state \citep{2007PhR...442..109L}.

Radial velocity (RV) and astrometric methods have been widely used to detect compact objects hidden in binaries \citep{2022NatAs...6.1085S,2022A&A...664A.159M,2022ApJ...938...78L,2022MNRAS.517.4005M,2022NatAs...6.1203Y,2024NatAs...8.1583W, 2024OJAp....7E..27E}, and are expected to provide a more complete picture of the compact object mass distribution.
By combining data from Large Sky Area Multi-Object Fiber Spectroscopic Telescope (LAMOST) and Apache Point Observatory Galactic Evolution Experiment (APOGEE), we have recently identified a single-lined spectroscopic binary system, {\it Gaia} ID 2546630699873483008 (R.A. = 2.00510$^\circ$, Decl. = 1.48284$^\circ$; hereafter J0008+0128), which exhibits a quite wide ($P_{\rm orb} \approx 1220$ days) but low-eccentricity ($e \lesssim 0.1$) orbit, consistent with the {\it Gaia} spectroscopic orbital fit in the nss\_two\_body\_orbit (NTBO) table.
Together with the astrometric solutions from {\it Gaia} DR2 and DR3, we determined that the unseen companion is a compact object with a mass of $1.8$--$2.8\ M_{\odot}$ (encompassing the conservative envelope of all mass estimates), placing it likely within the mass gap between NSs and BHs.

\subsection*{Results}

\subsubsection*{Spectral Observations and RV measurement}

We obtained sixteen infrared spectra from APOGEE and nine optical spectra from LAMOST, and performed two additional observations with the Nordic Optical Telescope (NOT) for J0008+0128 (see Supplement Materials).
The RVs for each spectrum were measured using the cross-correlation function (CCF) method.
For APOGEE spectra, we used the 15150--15700 \AA\ region. For LAMOST low-resolution spectra (LRS), we adopted the 5000--6000 \AA\ range, while for medium-resolution spectra (MRS), we used both the blue band (4950--5350 \AA) and red band (6350--6850 \AA). For NOT spectra, the wavelength range from 5500 \AA\ to 6800 \AA\ was employed.
Furthermore, we determined the RV zero point (RVZP) for each individual spectrum (see Supplement Materials).
Table \ref{rvdata.tab} lists the RVZP-corrected RV values, with uncertainties calculated as the square root of the sum of the squares of the measurement error and the RVZP error.

\subsubsection*{Stellar parameters of the visible star}  
\label{sec:stellar_parameters}

\begin{figure*}
    \center
    \includegraphics[width=0.98\textwidth]{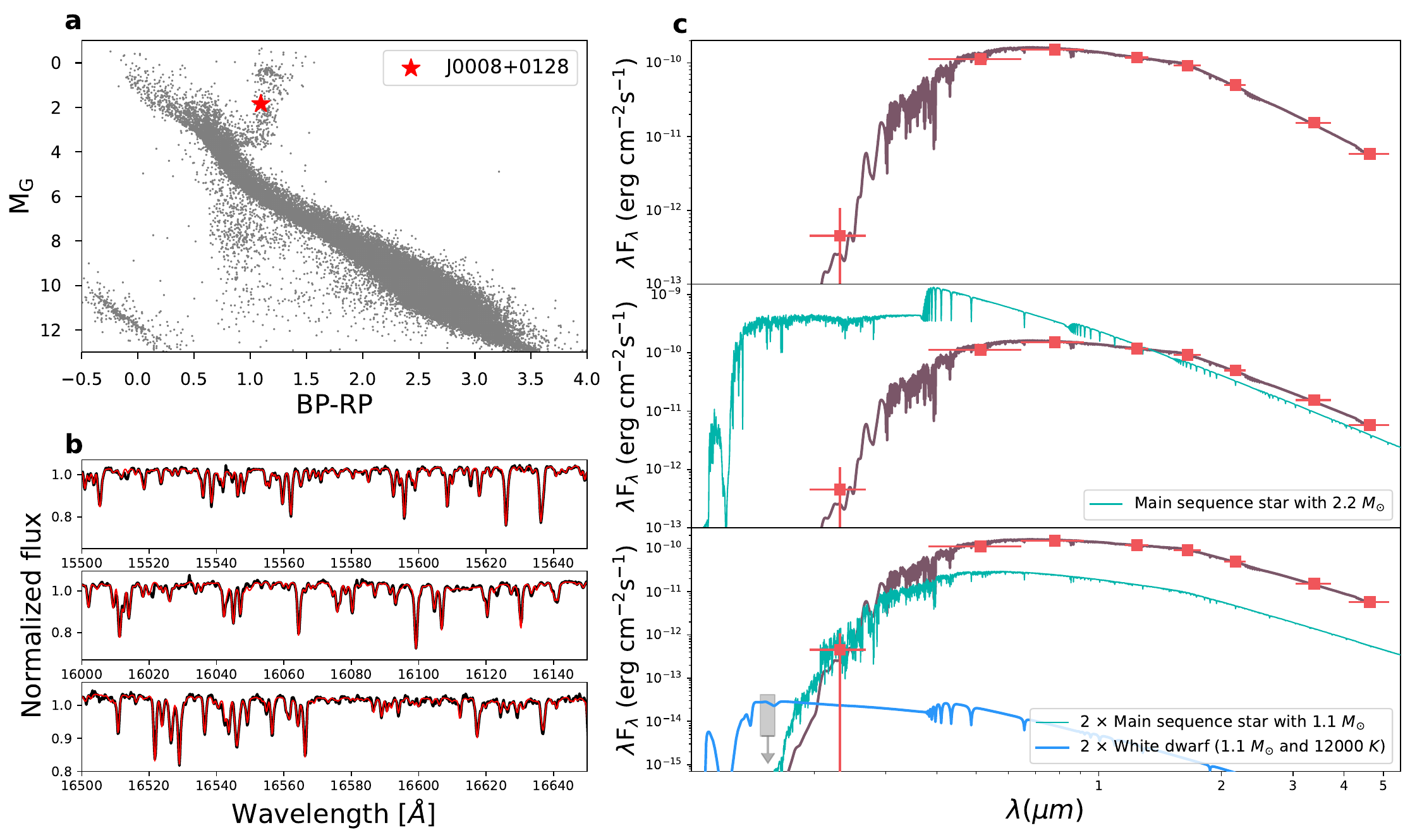}
    \caption{{\bf Spectroscopic and photometric information of J0008+0128.} {\bf Panel a:} Position of the visible star on the Hertzsprung–Russell diagram for J0008+0128. The gray points are plotted for a comparison, which are from {\it Gaia} EDR3 with distances $d <$ 100 pc, $G_{\rm mag}$ between 4--18 mag, and galactic latitudes $|b|$ $>$ 40 deg. 
    {\bf Panel b:} Comparison between the APOGEE spectrum (black line) and the best-fit template of the aspcapStar model (red line). 
    {\bf Panel c:} SED fitting of J0008+0128. Top: The red squares shows the observed multi-band magnitudes from {\it GALEX} (NUV), {\it Gaia} ($BP$ and $RP$), and {\it 2MASS} ($J$, $H$, and $K_{\rm S}$). The brown line is the best-fitting model returned by the {\it astroARIADNE} module. Middle: The cyan line plots the theoretical spectrum of a main sequence star with a mass of 2.2 $M_{\odot}$. Bottom: The cyan line represents the combined spectrum of two main sequence stars with a mass of 1.1 $M_{\odot}$, while the blue line is the combined spectrum of two WDs with mass of 1.1 $M_{\odot}$ and temperature of 12,000 K. The gray shaded region represents the upper limit in the {\it GALEX} FUV band.} 
    \label{fig1.fig}
\end{figure*}

In the case of wide binaries (especially including an invisible companion), the {\it Gaia} parallax may be inaccurate due to the assumption of the single-star model.
To obtain a more accurate distance, we conducted a joint analysis of the observed RV data and the astrometric data from {\it Gaia} DR2 and DR3 (see Materials and Methods).
As a result, the updated distance estimate of J0008+0128 is $1464.56^{+67.31}_{-61.64}$ pc, similar to the {\it Gaia} DR3 value ($=$1423$\pm$69 pc) \citep{2021AJ....161..147B}.
Its extinction is $E(B-V)=0.01^{+0.02}_{-0.01}$ from the Pan-STARRS DR1 dust map \citep{2015ApJ...810...25G}.
Figure \ref{fig1.fig} plots the position of the visible star on the Hertzsprung–Russell diagram.

The stellar atmospheric parameters of J0008+0128 have been estimated by both APOGEE and LAMOST pipelines. 
However, parameter estimates from different spectroscopic surveys suffer from systematic offsets, and their uncertainties include only random errors, resulting in an underestimated mass uncertainty for the giant star. By comparing with reference datasets, we estimated the systematic offsets and method-dependent errors. 
We then corrected the parameter values for these offsets and revised their errors by adding the method errors in quadrature to the random errors, which allows us to obtain a more realistic assessment of the parameters and their uncertainties (see Supplement Materials).
A summary of these results is provided in Table \ref{atmo_pars.tab}.
We also derived the radius and effective temperature of J0008+0128 via spectral energy distribution (SED) fitting (see Materials and Methods), adopting the APOGEE DR17 atmospheric parameters and the updated distance as priors.
The SED fitting yields $T_{\rm eff} = 4929^{+26}_{-22}$ K and $R_{1} = 5.39^{+0.24}_{-0.24}\ R_{\odot}$, with the best-fit result shown in Figure \ref{fig1.fig}.

We employed two methods to estimate the mass of the visible star in J0008+0128.
First, we derived the spectroscopic mass using the relation $M_1 = gR{_1}{^2}/G$ \citep{1687pnpm.book.....N}, combining the radius from SED fitting and $\log g$ from spectral analysis. 
We performed $10^5$ Monte Carlo simulations by sampling $R_1$ and log$g$, and adopted the 50th, 16th, and 84th percentiles of the resulting mass distribution as the best value and its uncertainties. 
The resulting spectroscopic mass ranges from 1.2 to 2.7 $M_\odot$ (encompassing the conservative envelope of all mass
estimates), with a best mass estimate of $1.67^{+0.15}_{-0.15}\ M_\odot$ based on Bayesian estimation (see Materials and Methods). 
Second, we estimated the evolutionary mass using the {\it isochrones} package \citep{2015ascl.soft03010M}, by fitting the MESA Isochrones and Stellar Tracks \citep{2016ApJS..222....8D} with the atmospheric parameters, multi-band photometry ({\it Gaia} and $2MASS$), \textit{{\it Gaia}} DR3 parallax, and extinction.
Adopting APOGEE/DR17 parameters as priors yields $M_1 = 1.45^{+0.13}_{-0.14}\,M_\odot$ and $t = 2.68^{+0.90}_{-0.54}$ Gyr, while using APOGEE/SDSS-V parameters gives $M_1 = 1.52^{+0.12}_{-0.12}\,M_\odot$ and $t = 2.40^{+0.64}_{-0.42}$ Gyr. These evolutionary masses are consistent with the spectroscopic estimates.

\subsubsection*{Keplerian orbit fitting}
\label{sec:keplerian_orbit}

\begin{figure*}
    \center
    \includegraphics[width=0.99\textwidth]{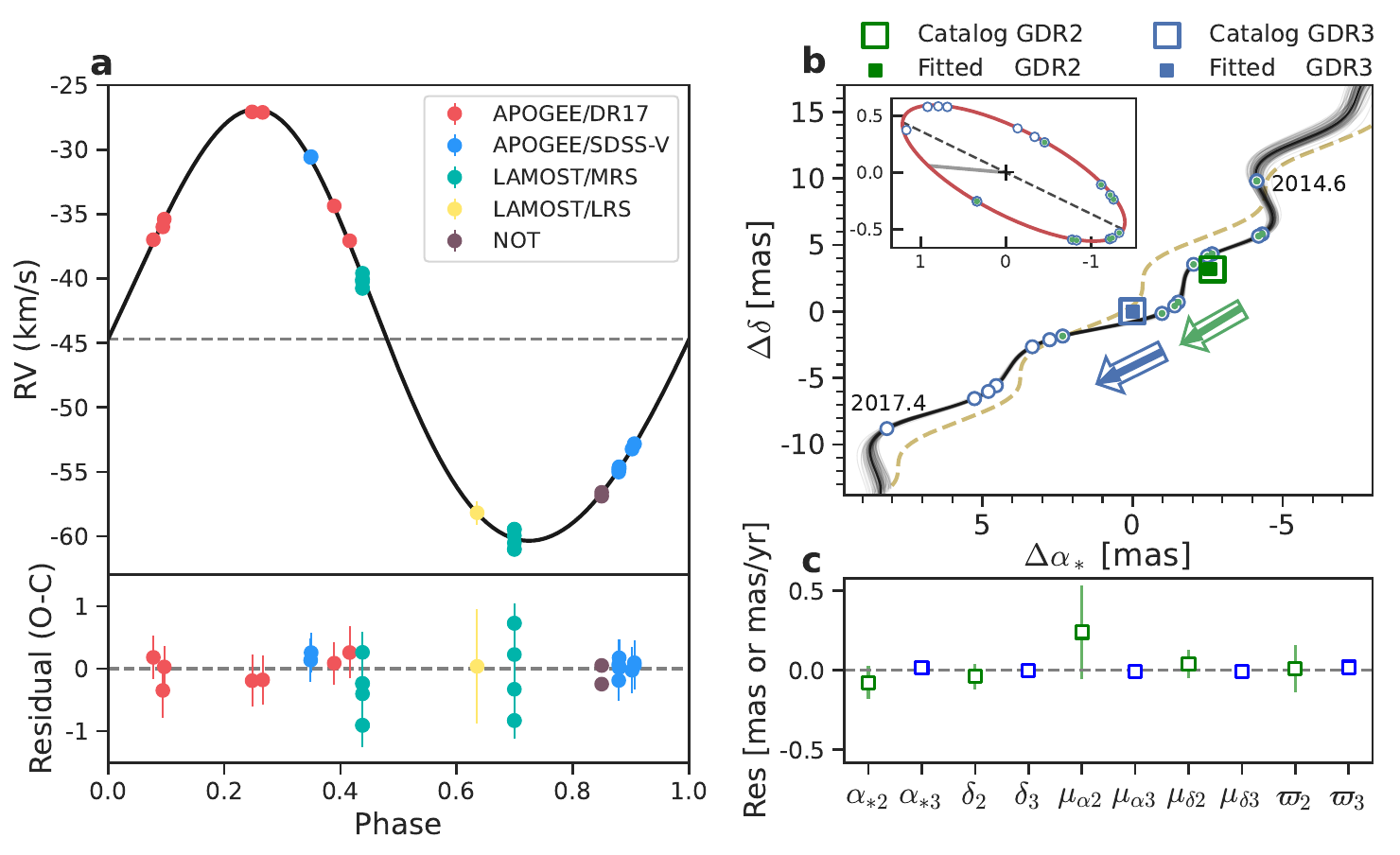}
    \caption{{\bf Orbital analysis for J0008+0128.} {\bf Panel a:} Folded RV data of J0008+0128 with a period of 1217.19 days. The subpanel shows the residuals between the observed RVs and the best-fitting model. {\bf Panel b:} Best-fitting astrometric solution for the \textit{Gaia} DR2 and DR3 data. The thick black curve represents the sky-projected motion of the photocentre (or the visible star), while the surrounding gray region shows representative trajectories randomly sampled from the MCMC chain, illustrating the uncertainty of the orbital solution. Filled circles and open circles mark the epochs at which the source crossed the \textit{Gaia} field of view during the DR2 and DR3 baselines, respectively, as predicted by GOST. The catalogued positions ($\alpha_{\ast}$, $\delta$) and proper motions ($\mu_{\alpha}$, $\mu_{\delta}$) from \textit{Gaia} are shown as open squares and arrows, whereas the corresponding best-fitting values are shown as filled symbols. For clarity, all positions are plotted relative to the \textit{Gaia} DR3 measurement. The yellow dashed curve indicates the motion expected from parallax and linear proper motion alone. The inset displays the sky-projected orbit of the star. The dashed black line denotes the line of nodes connecting the ascending and descending nodes, the ``+'' symbol marks the barycentre of the system, and the gray line connects the barycentre to the periapsis. {\bf Panel c:} Residuals between the \textit{Gaia} measurements ($\alpha_{\ast}$, $\delta$, $\mu_{\alpha}$, $\mu_{\delta}$ and parallax $\varpi$) and the best-fitting five-parameter astrometric model. The subscripts ``2'' and ``3'' correspond to the DR2 and DR3 epochs, respectively.}
    \label{fig2.fig}
\end{figure*}

We first performed RV fitting using {\it The Joker} (see Materials and Methods) and obtained results consistent with those from the {\it Gaia} DR3 NTBO table (Table \ref{keplerorbit.tab}).
The best-fit orbital parameters are:
period $P = 1217.19^{+0.58}_{-0.51}$\,days, 
eccentricity $e = 0.08^{+0.01}_{-0.01}$, 
argument of periastron $\omega = 0.62^{+0.02}_{-0.03}$, 
mean anomaly $M_0 = 0.57^{+0.02}_{-0.02}$, 
semi-amplitude $K_1 = 16.75^{+0.03}_{-0.03}$\,km/s, 
and systemic velocity $\nu_0 = -44.70^{+0.02}_{-0.02}$\,km/s (Figure \ref{fig2.fig}). 
This yields a binary mass function of $f(M_2) = 0.587^{+0.004}_{-0.004}\,M_\odot$. 
Assuming the inclination angle $i = 90^\circ$ and using the spectroscopic mass estimates of the visible star, we determined that the minimum mass of the invisible companion is 1.7--2.5 $M_{\odot}$.

To derive the orbital inclination, we then conducted a joint RV and astrometric fit using {\it Gaia} DR2 and DR3 data (see Materials and Methods). 
Using the spectroscopic masses of the visible star as priors, we derived detailed orbital parameters and the component masses (Table \ref{table1.tab}). 
The best-fit model is shown in Figure \ref{fig2.fig}.
The inclination is estimated at $i \approx 73^\circ\pm11^\circ$, yielding a companion mass of 1.8--2.8 $M_{\odot}$ (encompassing the conservative envelope of all mass estimates).
Using a Bayesian estimation approach, we obtained a best estimate of $2.20^{+0.14}_{-0.13}\ M_{\odot}$ for the companion (see Supplement Materials). 
If the companion is a compact object, it would likely be a supermassive NS. 

Additionally, no periodic signals were detected from multi-band light curves; however, comparison between the observed photometric amplitude (from TESS light curves) and theoretical prediction excludes orbital periods shorter than $\approx$30 days (Figure \ref{m2_inclination.fig}).

\subsection*{Discussion}

\subsubsection*{Whether the visible component is a stripped star?}  
\label{sec:stripped_star}

The mass of the visible star may be overestimated if the binary system were currently undergoing, or had previously experienced, mass transfer.
However, such scenarios can be ruled out for J0008+0128 based on spectroscopic and photometric analyses. 
First, using the measured orbital parameters and mass ratio, we derived that the Roche lobe filling factor of the visible star is only $\approx$2\% (see Materials and Methods), indicating no ongoing mass transfer. 
This is consistent with the presence of $H_{\alpha}$ in absorption rather than emission, which suggests the absence of an accretion disk.
Second, if the visible star were a stripped star, its large Roche lobe radius ($\approx 270\,R_\odot$) would imply a minimum mass of about $0.6\,M_\odot$ \citep{1995MNRAS.273..731R}, leading to a companion of approximately $1.3\,M_\odot$. 
If such a companion were a normal star, it would contribute detectable flux in the $BP$, $G$, and $RP$ bands, with predicted flux ratios of 0.29, 0.24, and 0.19, respectively. 
A partially stripped visible star (e.g., with remaining mass of 0.8, 1.0, or 1.2 $M_{\odot}$) can likewise be ruled out by the same method.
However, the observed spectrum of J0008+0128 matches well with a single-star template of similar parameters, showing no evidence of additional spectral features from any other component.
Third, if the visible star were a stripped star in a post–mass-transfer phase, its small radius-to–Roche-lobe ratio ($\approx$2\%) would imply an effective temperature exceeding 12000 K \citep{2017MNRAS.467.1874C,2018A&A...615A..78G}, far higher than observed.
We therefore conclude that the visible star in J0008+0128 has not undergone notable mass transfer, and that the mass measurements (See Materials and Methods) are robust.

\subsubsection*{The nature of the dark component} 
\label{sec:nature_of_companion}

\begin{figure*}
    \center
    \includegraphics[width=1\textwidth]{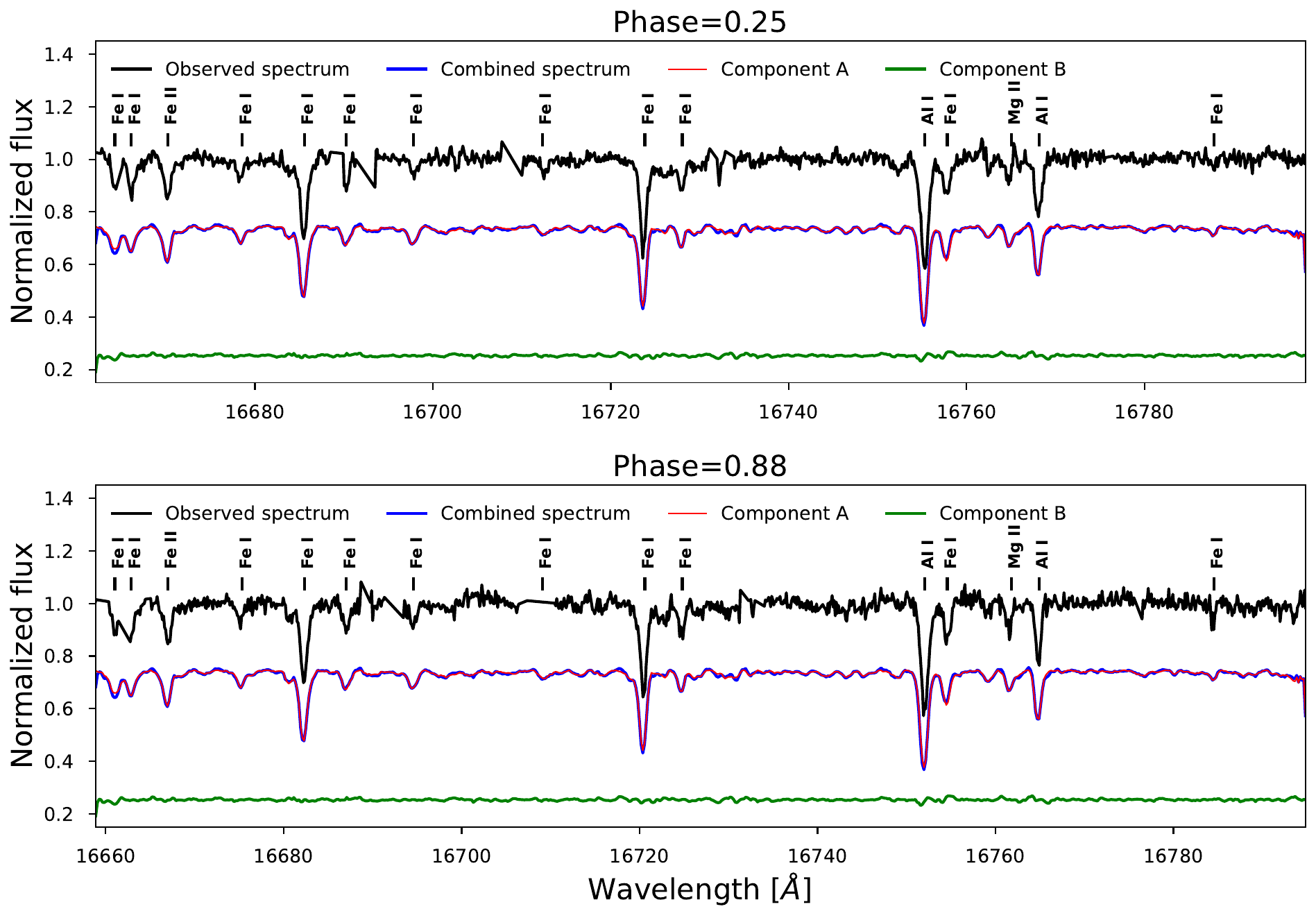}
    \caption{{\bf Spectral disentangling results at phases of large redshift and blueshift.} The companion is assumed to be a normal star with a mass of 2.2 $M_{\odot}$. Black lines indicate the observed spectra. 
    Red and green lines represent the disentangled components, while blue lines shows the combined spectra.}
    \label{fig3.fig}
\end{figure*}

The unseen companion is unlikely to be a single or pair of main-sequence stars.
A normal  main-sequence star with a mass of 1.8--2.8 $M_{\odot}$ has a luminosity about 11--81 $L_{\odot}$, which would dominate or at least be a strong component of the SED and spectrum, given that the observed luminosity of J0008+0128 is $\approx$16 $L_{\odot}$.
However, no such signatures are present in either the SED or the observed spectra (Figure \ref{fig1.fig}). 
Similarly, a companion consisting of one or two Sun-like stars would produce detectable flux excess in optical bands (particularly in $U$, $B$, and $BP$), contrary to the observation. 
Most importantly, spectral disentangling confirmed that there is no light contribution from any additional stellar component (see Materials and Methods).
Figure \ref{fig3.fig} shows the disentangled profiles obtained under the assumption that the companion is a normal star of mass 2.2 $M_{\odot}$, plotted at phases of large redshift and blueshift.
Therefore, if the companion is a single compact object, it would be a supermassive NS. 
However, a binary compact-object companion, such as a white dwarf (WD) and an NS, cannot be totally excluded with current photometric and spectroscopic data.
In addition, a low-mass black hole (BH) is also a viable alternative, given that the upper mass limit of J0008+0128 approaches 3 $M_{\odot}$, which places it in the mass gap between NSs and BHs, analogous to the massive companion of PSR J0514-4002E.

\subsubsection*{X-ray and radio follow-up observations}  
\label{sec:Xray_fast}

We performed X-ray follow-up observation of J0008+0128 using the Einstein Probe (see Supplement Materials). 
No X-ray source was detected within 2 arcmin of the target at the $3\sigma$ significance level. 
The $0.5$--$10$\,keV X-ray luminosity is constrained to be below $8.8 \times 10^{30}\ \mathrm{erg\ s^{-1}}$ for a power-law index of $\Gamma = 1.5$ and $6.0 \times 10^{30}\ \mathrm{erg\ s^{-1}}$ for $\Gamma = 2.0$. 

We assumed that J0008+0128 hosts an NS. To check whether the giant star plus NS scenario is consistent with the observed X-ray upper limit, we examined possible sources of X-ray emission.
First, assuming wind accretion onto the NS, we estimated the X-ray luminosity using the Bondi-Hoyle-Lyttleton theory, obtaining only $\sim$10$^{26}$--10$^{27}$ $\mathrm{erg\ s^{-1}}$.
Second, considering thermal emission from the NS surface, a 2.2 $M_{\odot}$ NS would cool rapidly, with its X-ray luminosity dropping below $5\times10^{30}$ $\mathrm{erg\ s^{-1}}$ within 10$^{3}$--10$^{5}$ years. 
Third, we considered coronal emission from the giant star. Its very low NUV flux, extremely weak Ca II H\&K emission, and lack of $H_{\alpha}$ emission all support very weak magnetic activity, and hence low X-ray luminosity.
Therefore, the observed X-ray upper limit is consistent with the scenario of a giant star-NS binary.

We also applied for radio observation using the Five-hundred-meter Aperture Spherical Telescope (FAST) (see Supplement Materials). 
No persistent or pulsed radio emission was detected by FAST, with a sensitivity limit of $\sim 2\ \mu$Jy in the frequency range from 1.05 to 1.45 GHz. 
The orbital and stellar parameters of J0008+0128 are consistent with a non-recycled pulsar in a wide orbit with a non-degenerate companion.

\subsubsection*{Comparison with other known NSs}
\label{sec:comparison}

\begin{figure*}
    \center
    \includegraphics[width=1\textwidth]{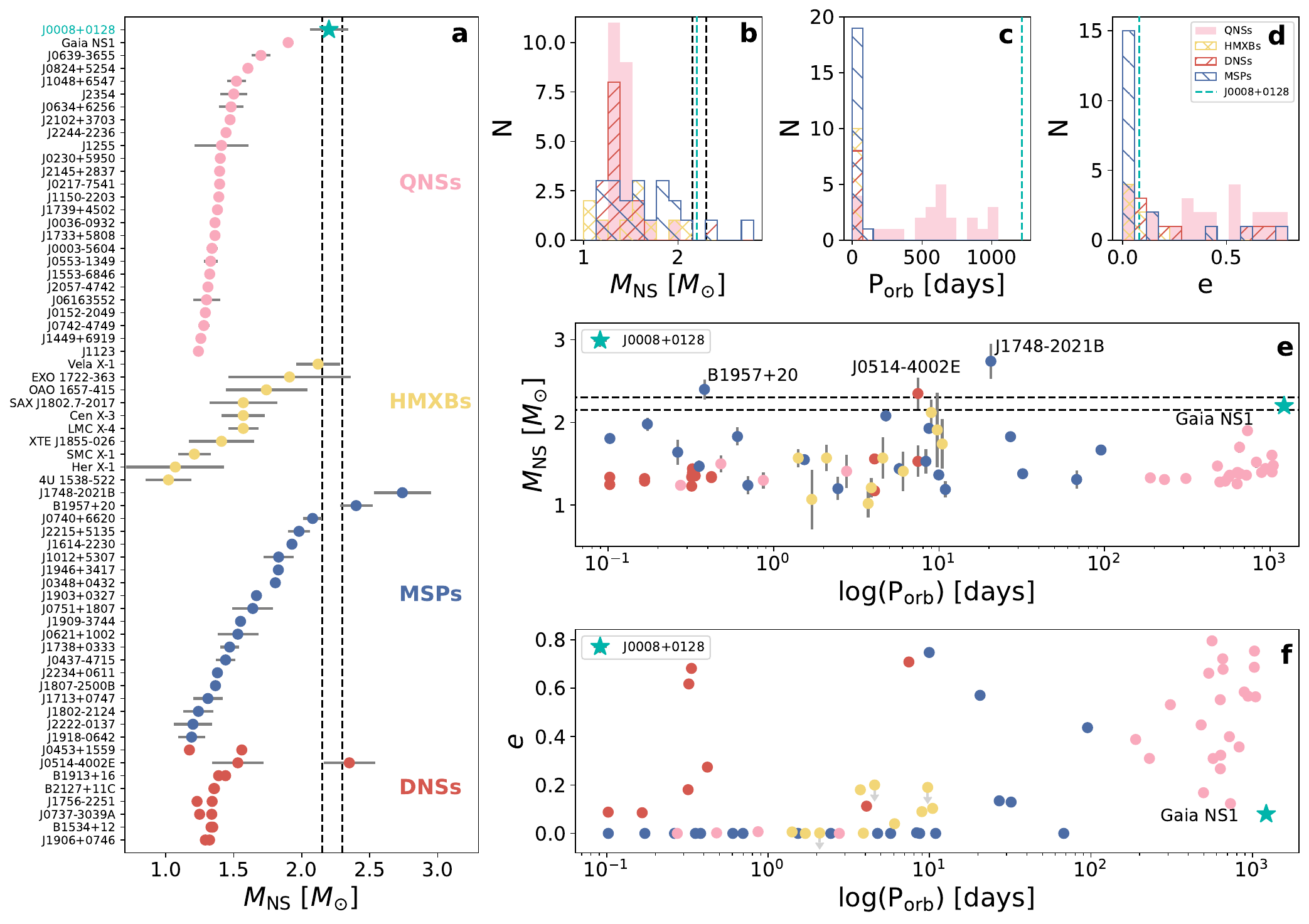}
    \caption{{\bf Comparison of J0008+0128 to other NS systems.} {\bf Panel a:} The masses of NSs in different systems. Pink, yellow, blue, and red symbols denote QNSs, HMXBs, MSPs, and DNSs, respectively. The same color scheme is adopted for these NS subclasses in all other subpanels. {\bf Panel b:} Mass distribution for J0008+0128 and other NS systems. {\bf Panel c:} Orbital period distribution. {\bf Panel d:} Eccentricity distribution. {\bf Panel e:} The orbital period vs. the masses of NSs. {\bf Panel f:} The orbital period vs. eccentricity. 
    The cyan star marks the location of J0008+0128 with its best-fit parameters. The cyan dashed lines highlight the position of J0008+0128. The black dashed lines denote the  maximum mass range (2.15$\sim$2.3 $M_{\odot}$) for non-rotating or slowly rotating NSs based on the equation of state \citep{2017ApJ...850L..19M,2018ApJ...852L..25R,2019PhRvD.100b3015S}.}
    \label{fig4.fig}
\end{figure*}

We compared J0008+0128 with different types of NS systems, such as double NSs (DNSs), millisecond pulsars (MSPs), high-mass X-ray binaries (HMXBs), and quiescent NSs (QNSs). The orbital parameters and component masses of these systems are summarized in Table \ref{other_ns.tab}.
Figure \ref{fig4.fig} shows the distributions of their systematic parameters.
HMXBs and MSPs typically host more massive NSs and exhibit shorter orbital periods (from hours to tens of days). 
These large NS masses are likely the result of accretion from their companions, implying that their present-day masses exceed their birth masses.
By contrast, DNSs and QNSs tend to contain less massive NSs and have longer orbital periods, suggesting negligible mass transfer during their evolution.
In comparison, J0008+0128 has the longest orbital period and likely hosts one of the most massive NSs.
It shares similarities with {\it Gaia} NS1, a wide binary ($P=$639 days) consisting of a K-type star and a 1.9~$M_{\odot}$ NS \citep{2024OJAp....7E..27E}.
Generally, those most massive NSs are consistent with the prediction of the maximum NS mass between $\sim$2.15 and $\sim$2.30 $M_{\odot}$ \citep{2017ApJ...850L..19M,2018ApJ...852L..25R,2019PhRvD.100b3015S}.

\subsubsection*{Formation path: Isolated binary evolution}
\label{sec:formation_binary}

Binaries can, in principle, form via tidal capture, in which the inbound companion raises a tide on the target star that allows sufficient energy loss and could result in a bound orbit. However, in the Galactic field, such close encounters are very rare. 
Moreover, the resulting orbit is highly eccentric. Given the long orbital period of J0008+0128, tidal circularization could not have acted to give the binary its comparatively low present-day eccentricity.

We then examined the population of solar-metallicity binaries simulated with the POSYDON (version 1) binary population synthesis framework \citep{Fragos2023,Andrews2024}, to assess the likelihood of forming such a system from an isolated intrinsic binary.
We performed a systematic search over the full binary population, scanning the entire evolutionary history of all simulated systems 
to identify epochs with properties similar to J0008+0128. Specifically, we required orbital periods in the range $1000$--$1500\,\mathrm{d}$, eccentricities $e < 0.2$, primary masses between 1.8 and $3.0 \,M_\odot$, and secondary masses between 1.0 and $2.9\,M_\odot$ (corresponding to the visible star) (See Supplement Materials).
Among $\sim 3\times10^{5}$ simulated binaries from POSYDON, we found no system enters the parameter region defined above during its evolutionary history, meaning that none reproduces physical and orbital properties similar to J0008+0128.
The scarcity of J0008+0128--like systems is unlikely to be driven solely by limited POSYDON sampling, and instead points to a strong evolutionary bias within isolated binary evolution. Nonetheless, the formation of such a system is not precluded in principle, and may require a combination of extreme but physically plausible parameters, such as an unusually efficient common envelope (CE) ejection during the evolution of the NS progenitor, a high NS birth mass, and minimal supernova kick.

Assuming an initial circular orbit ($e=0$) and symmetric mass ejection of $\Delta M$ from the NS progenitor, the resulting eccentricity can be expressed as $e = \Delta M / (M_{\rm G} + M_{\rm NS})$. With the measured eccentricity $e = 0.09\pm0.01$, the ejected mass $\Delta M$ ($= M_{\rm He} - M_{\rm NS}$) must be less than 10\% of the remaining mass, implying a total mass loss of roughly 0.4 $M_{\odot}$.
Consequently, the He star progenitor likely had a mass no larger than $\approx$2.6 $M_{\odot}$.
This upper limit, however, is notably below the predictions of recent stellar-evolution models, which shows that a 2.2 $M_{\odot}$ NS may be produced from stripped He stars with masses of 6--8 $M_{\odot}$, originating from main‑sequence stars of 19--21 $M_{\odot}$ at solar metallicity \citep{2021ApJ...916L...5V}.
This discrepancy calls for extreme stripping of the He star progenitor of J0008+0128.
Given the high mass ratio between the NS progenitor and the visible giant, mass transfer is expected to be unstable, triggering a CE phase.
Explaining the survival of the binary after CE ejection calls for a very high CE ejection efficiency, implying either a low envelope binding energy or a substantial external energy input. This problem is also encountered in several wide binaries hosting BHs or NSs, such as {\it Gaia} BH1 \citep{2023MNRAS.518.1057E} and BH2 \citep{2023MNRAS.521.4323E}, G3425 \citep{2024NatAs...8.1583W}, and {\it Gaia} NS1 \citep{2024OJAp....7E..27E}.
Simultaneously, the natal kick during the supernova explosion must have been small, consistent with the limited mass loss. Similar to the case of {\it Gaia} NS1 \citep{2024OJAp....7E..27E}, a kick velocity below about 20 km/s would be required.

\subsubsection*{Formation path: Triple system evolution}
\label{sec:formation_triple}

Triple systems are common in the universe, especially among massive early-type (B/O) primaries \citep{2017ApJS..230...15M}. 
Within this scenario, J0008+0128 may have originated as a hierarchical triple system, with the observed giant star's progenitor as the outer tertiary and an inner binary composed of two massive stars.
Three-body dynamics, combined with stellar evolution, often drive strong interactions that can lead to a merger of the inner binary \citep{2025ApJ...978...47S}.
For example, after the more massive star collapses into an NS and its companion enters into a CE phase, the NS may spiral into the companion’s core before the envelope is ejected. This could create a massive Thorne-\.{Z}ytkow object \citep{1995MNRAS.274..485P} and ultimately lead to the formation of the present NS via accretion.
Alternatively, the CE might have been successfully ejected before the merger, leaving the central unseen object as the product of two merged NSs.
During the long-term evolution of the system, Kozai–Lidov oscillations \citep{1962AJ.....67..591K,1962P&SS....9..719L} may help shrink the orbit of the inner binary and accelerate its eventual merger.

Such triple-system channels have been proposed to explain the 2.09--2.71 $M_{\odot}$ companion in PSR J0514-4002E \citep{2024Sci...383..275B} and the 2.6 $M_{\odot}$ component in GW190814 \citep{2021MNRAS.500.1817L}.
The inferred mass of the unseen compact object in J0008+0128 is also comparable both to the remnant mass of GW170817 ($\lesssim$2.3 $M_{\odot}$) \citep{2019PhRvD.100b3015S} and to the total masses of known DNSs (2.5--2.9 $M_{\odot}$) \citep{2019ApJ...876...18F}, suggesting a likely origin as the merger product of two NSs.
An NS+WD merger is also possible, such as a massive NS with a low-mass WD (e.g., 1.8 $M_{\odot}$ $+$ 0.4 $M_{\odot}$), or a typical NS with a normal WD (e.g., 1.5 $M_{\odot}$ $+$ 0.7 $M_{\odot}$). 
One important point is that the merger of two NSs or an NS–WD binary is unlikely to produce a large kick, which thus helps to produce a low‑eccentricity orbit.
In addition, the existence of the hierarchical triple system PSR J0337+1715, an inner NS+WD binary orbited by an outer WD \citep{2014Natur.505..520R, 2014ApJ...781L..13T}, demonstrates that similar formation pathways are physically feasible.
The formation of PSR J0337+1715 may involve a three‑body CE episode, one or more stable mass‑transfer stages, and even an accretion‑induced collapse \citep{2014ApJ...781L..13T, 2015MNRAS.450.1716S}.

It is also possible that the central object remains an unresolved compact binary, containing two lower-mass compact objects, such as two NSs, or an NS and a WD. 
Assuming such an inner binary with component masses of 1 $M_{\odot}$ each, we used the {\it rebound} integrator \citep{rebound,reboundias15} to simulate the RV signal induced on the visible star for a range of inner orbital periods. The simulations indicate that such inner binaries with orbital periods longer than about 20–-30 days can be roughly ruled out (Figure \ref{2n_3n.fig}).
Further short-cadence spectroscopic follow-up is essential to test this hypothesis by searching for short-period RV modulations, although detecting such signals is extremely challenging.

\subsection*{Materials and Methods}\label{sect:methods}

\subsubsection*{Spectral energy distribution fitting}
\label{sec:sed_fitting}

Spectral energy distribution (SED) fitting provides a method to estimate stellar radii and effective temperatures using photometric data. 
On the other hand, SED fitting can also be used to roughly determine the presence of additional components in the observed spectrum.
We estimated the stellar parameters of the visible star using the {\it astroARIADNE} Python module, which can automatically fit the broadband photometry.
The multi-band magnitudes include the NUV magnitude from the Galaxy Evolution Explorer ({\it GALEX}), the $BP$ and $RP$ magnitudes from {\it Gaia},  $J$, $H$, and $K_{\rm S}$ magnitudes from the Two Micron All Sky Survey ({\it 2MASS}), $W$1 and $W$2 magnitudes from the Wide-field Infrared Survey Explorer ({\it WISE}).
Using the updated distance of $1464.56^{+67.31}_{-61.64}$ pc (See following Materials and Methods), we derived an extinction of $E(B-V)=0.01^{+0.02}_{-0.01}$ from Pan-STARRS DR1 dust map \citep{2015ApJ...810...25G}. This is consistent with the small total Galactic extinction in this direction, with a value of $E(B-V)=0.027$ from SFD dust map \citep{1998ApJ...500..525S}.
With the APOGEE DR17 atmospheric parameters, distance, and extinction adopted as priors, the stellar parameters of the visible star from SED fitting are $T_{\rm eff} = 4929^{+26}_{-22}$ K, log$g$ $=$ $3.13^{+0.08}_{-0.04}$, [Fe/H] $= -0.22^{+0.03}_{-0.08}$, and $R_{1}=5.39^{+0.24}_{-0.24} R_{\odot}$, consistent with spectroscopic estimations.
Figure \ref{fig1.fig} shows the best-fitting in the {\it astroARIADNE}.

Additionally, we calculated the upper limit magnitude of FUV band using {\it GALEX} images.
We queried the {\it GALEX} GR6+7 database through CasJobs and obtained 7 observations of the source in both the NUV and FUV bands. 
Following the photometric procedure described in Li et al. \citep{2025ApJS..276...29L}, we performed aperture photometry on the {\it GALEX} images using the reduction pipeline {\it photutils} and applied point-spread-function corrections to account for the finite aperture. In the NUV band, we obtained magnitudes in the range of 20.0--20.2 mag, which agree very well with the catalog values, confirming the reliability of our photometry process. In the FUV band, the source is too faint to be detected; we therefore provide an upper-limit range of 25.2--27.5 mag.

\subsubsection*{Mass determination}
\label{sec:mass_determination}

The stellar radii ($R$) and the surface gravity ($g$) can be used to estimate the spectroscopic mass using $M=gR^2/G$.
We calculated the spectroscopic mass of the visible star using the radius derived from SED fitting and log$g$ from spectral analysis.
A total of 10$^5$ Gaussian sampling iterations of $R$ and log$g$ are adopted.
The 50th, 16th, and 84th percentile of the mass estimates were adopted as the final mass and its lower and upper limits, respectively.
Finally, we determined the spectroscopic masses of J0008+0128 as 
$1.46^{+0.29}_{-0.24}\ M_{\odot}$,
$1.64^{+0.29}_{-0.25}\ M_{\odot}$, 
$1.97^{+0.44}_{-0.36}\ M_{\odot}$,
$2.11^{+0.59}_{-0.46}\ M_{\odot}$,
and $1.68^{+0.47}_{-0.37}\ M_{\odot}$
for the stellar parameters derived from  APOGEE/DR17 (ASPCAP), APOGEE/SDSS-V (ASPCAP), LAMOST/LRS (LASP), LAMOST/LRS (DD-Payne), and LAMOST/MRS (LASP), respectively.
A Bayesian estimation was performed based on the spectroscopic masses, yielding a best-fit value of $1.67^{+0.15}_{-0.15}\ M_{\odot}$ for the visible star (See Supplementary Materials).

Stellar evolution models also provide a way to estimate the stellar mass for a single star.
This mass is called evolutionary mass.
However, the evolutionary mass derived from stellar evolution models may be inaccurate \citep{2022MNRAS.516.5945J,2022MNRAS.512.5620E,2024ApJ...964..101Z} for stars in binaries undergoing mass transfer (such as stellar winds and accretion), especially for binaries with strong accretion processes.
To estimate the evolutionary mass of the visible star, we used the {\it isochrones} Python module \citep{2015ascl.soft03010M}, which fits the photometric or spectroscopic parameters with MIST models and returns observed and physical parameters.
We took the atmospheric parameters ($T_{\rm eff}$, log$g$, [Fe/H]) derived from APOGEE,  multi-band magnitudes from {\it Gaia} and {\it 2MASS}, {\it Gaia} DR3 parallax, and extinction as the input priors. 
Using APOGEE/DR17 parameters, we derived the mass, radius, and age as $M=1.45^{+0.13}_{-0.14}\ M_{\odot}$, $R=5.34^{+0.04}_{-0.04}\ R_{\odot}$, and $t=2.68^{+0.90}_{-0.54}$ Gyr, respectively, while APOGEE/SDSS-V yields $M=1.52^{+0.12}_{-0.12}\ M_{\odot}$, $R=5.35^{+0.04}_{-0.04}\ R_{\odot}$, and $t=2.40^{+0.64}_{-0.42}$ Gyr.
The spectroscopic mass is consistent with the evolutionary mass for the visible star.

\subsubsection*{Radial velocity fitting} 
\label{sec:rv_fitting}

 We performed RV fitting using the custom Markov Chain Monte Carlo (MCMC) sampler {\it The Joker} \citep{2017ApJ...837...20P}.
We first ran the fit over several period ranges (e.g., 0 to 200 days, 200 to 500 days, 500 to 1000 days, and 1000 to 1300 days), which yielded a robust solution only for an orbital period of about 1220 days.
We then carried out a more refined fit over the narrower range of 1150--1250 days.
The fitted orbital parameters are: 
period $P=1217.19^{+0.58}_{-0.51}$ days,
eccentricity $e = 0.08^{+0.01}_{-0.01}$, 
argument of the periastron $\omega =0.62^{+0.02}_{-0.03}$, 
mean anomaly at the first exposure $M_{\rm 0} =0.57^{+0.02}_{-0.02}$, 
semi-amplitude $K_{\rm 1} =16.75^{+0.03}_{-0.03}$ km/s, 
and systematic RV $\nu_{\rm 0} =-44.70^{+0.02}_{-0.02}$ km/s,
in good agreement with the results from the {\it Gaia} DR3 NTBO table  \citep{2023A&A...674A..34G}.
Figure \ref{fig2.fig} shows the folded RV data and the RV curve obtained from {\it The Joker}.

The mass function $f(M)$ were calculated as follows:
\begin{equation}
    f(M_{2}) = \frac{M_{2} \, \textrm{sin}^3 i} {(1+q)^{2}} = \frac{P \, K_{1}^{3} \, (1-e^2)^{3/2}}{2\pi G},
\label{mf}
\end{equation}
\noindent
where $M_{2}$ is the mass of the companion in binary system, $q = M_{1}/M_{2}$ is the mass ratio, and $i$ is the system inclination.
The mass function of J0008+0128 derived from {\it The Joker} is $f(M_{2}) = 0.587^{+0.004}_{-0.004}\ M_{\odot}$, suggesting a companion with a minimum mass ($i=90^{\circ}$) of $1.86^{+0.15}_{-0.17}\ M_{\odot}$, $1.97^{+0.15}_{-0.16}\ M_{\odot}$, $2.15^{+0.21}_{-0.22}\ M_{\odot}$, $2.23^{+0.26}_{-0.29}\ M_{\odot}$, and $1.99^{+0.22}_{-0.25}\  M_{\odot}$, corresponding to the five spectroscopic mass measurements of the visible star.

\subsubsection*{Joint fitting of Radial velocity and astrometry}
\label{sec:RV_and_astrometry}

{\it Gaia} DR2 and DR3 (hereafter GDR2/3) generally provide five-parameter astrometric solutions under the assumption of a single-star model, including right ascension ($\alpha$), declination ($\delta$), parallax ($\varpi$), and proper motions in right ascension and declination ($\mu_\alpha$ and $\mu_\delta$).
For unresolved binaries with notable orbital motion, a single-star model may fail to fully account for the source motion, leading to inaccurate parallax estimate. Simultaneously, because the reflex motion induced by the companion is absorbed into the astrometric parameters, the {\it Gaia} solutions can be used to infer the binary's orbital properties.
Therefore, here we performed a joint orbital fitting with RV and {\it Gaia} astrometry using the similar method of Feng et al. \citep{Feng2023MNRAS} and Xiao et al. \citep{Xiao2024MNRAS}. 
This method has been used to constrain the full orbit of exoplanets, brown dwarfs and other dark companions (e.g., \citep{Xiao2024MNRAS,2024NatAs...8.1583W}).

For the RV model, the logarithmic likelihood is
\begin{equation}
\label{rvmodel}
  \mathcal{L}_{\rm rv}=\prod\limits_{i=1}^{N_{\rm rv}}\frac{1}{\sqrt{2\pi(\sigma_{i}^2+J_{{\rm rv}}^2)}}{\rm exp}\left(-\frac{(v_{i}-\hat{v}_{i})^2}{2(\sigma_{i}^2+J_{{\rm rv}}^2)}    \right)~,
\end{equation}

where $J_{\rm rv}$ is the excess RV noise (called "RV Jitter"), $\hat{v}_i$ is the model-predicted RV and $v_i\equiv v(t_i)$ is the measured RV at epoch $t_i$.

 For the astrometric model, we first construct the target system barycenter (TSB) at the {\it Gaia} DR3 reference epoch.
Given that the {\it Gaia} astrometric time series data is currently unavailable,  we simulated both DR2 and DR3 along-scan (AL) abscissae using the {\it Gaia} Observation Forecast Tool (\url{https://Gaia.esac.esa.int/gost/}) (GOST). This simulation is achieved by adding the stellar reflex motion and parallax motion into the linear motion of TSB. The synthetic {\it Gaia} abscissa $\eta_i$ at each GOST epoch $t_i$ is described as below:
\begin{eqnarray}
  \eta_i=(\alpha_{*i}+\Delta\alpha_{*i}^r)\sin\psi_i+(\delta_i+\Delta\delta_i^r)\cos\psi_i+\varpi_i f_i^{\rm AL}, 
\end{eqnarray}
where ($\alpha_{*i}$, $\delta_i$) is the linear motion of TSB, ($\Delta\alpha_{*i}^r$, $\Delta\delta_i^r$) is the reflex motion induced by a companion, $f_i^{\rm AL}$ is the parallax factor at epoch $t_i$ and $\psi_i$ is the scan angle. $\alpha_*$ is defined as $\alpha_*=\alpha\cos\delta$.

We then fitted the synthetic abscissa measurements using a five-parameter linear model analogous to the {\it Gaia} single-star astrometric model, and compared the resulting parameters with the {\it Gaia} catalogue astrometry.
The likelihood of {\it Gaia} astrometry can be written as follows:
\begin{equation} 
\begin{split}
\mathcal{L}_{\rm Gaia} &= \frac{1}{\sqrt{(2\pi)^5|\Sigma_k(S^2)|}} \times \\
&\left( {\rm exp}\left(-\frac{1}{2}(\hat\vec{\iota}_{\rm GDR2}-\vec{\iota}_{\rm GDR2})^T[\Sigma_k(S^2)]^{-1}
(\hat\vec{\iota}_{\rm GDR2}-\vec{\iota}_{\rm GDR2})\right) \right. \\
&\left.+{\rm exp}\left(-\frac{1}{2}(\hat\vec{\iota}_{\rm GDR3}-\vec{\iota}_{\rm GDR3})^T[\Sigma_k(S^2)]^{-1}
(\hat\vec{\iota}_{\rm GDR3}-\vec{\iota}_{\rm GDR3})\right)\right) 
~.
\end{split}
\end{equation}
The subscripts "GDR2" and "GDR3" indicate {\it Gaia} DR2 and {\it Gaia} DR3, respectively. $\hat\vec{\iota}$ and $\vec{\iota}$ represent the fitted and observed astrometric parameters, and $\Sigma$ means the covariances of astrometric parameters. $S$ is the error inflation factor for {\it Gaia} astrometry. With the total likelihood ($\mathcal{L}=\mathcal{L}_{\rm rv}\cdot\mathcal{L}_{\rm Gaia}$), we finally derive the orbital solution by sampling the posterior via the parallel-tempering MCMC sampler \texttt{ptemcee} \citep{Vousden2016}. More details on the methodology can be found in Appendix B of Xiao et al. \citep{Xiao2024MNRAS}.

Although the observational baseline of {\it Gaia} ($\sim$ 1000 days) only covers about 82\% of J0008+0128’s orbital period, it provides sufficient phase sampling to significantly reduce the uncertainties in the derived orbital solutions.
Furthermore, the astrometric solutions in {\it Gaia} DR3 show high precision and small uncertainties, allowing us to narrow down the possible range of the astrometric and orbital solutions.

First, we used both RV data and astrometric parameters from {\it Gaia} DR2 and DR3 to derive the astrometric solutions. A rough spectroscopic mass of the visible star (estimated from the {\it Gaia} DR3 distance) was adopted as a prior, but the astrometric solutions are fortunately insensitive to the stellar mass. This step yielded a distance of $1464.56^{+67.31}_{-61.64}$ pc, which differs notably from the {\it Gaia} DR2 distance of $1185\pm205$ pc \citep{2018AJ....156...58B}, but is in much better agreement with the {\it Gaia} DR3 distance of $1423\pm69$ pc \citep{2021AJ....161..147B}. This is not unexpected, as the different observation baselines of DR2 and DR3 can lead to measurable differences in their astrometric solutions for binaries containing an unseen object, similar to the cases of {\it Gaia} BH2 \citep{2023MNRAS.521.4323E} and G3425 \citep{2024NatAs...8.1583W}.
Second, we used the updated distance for SED fitting to obtain an accurate radius and thus a refined spectroscopic mass.
Third, we performed the joint fitting again, using this updated mass of the visible star, together with the RV data and astrometric parameters from {\it Gaia} DR2 and DR3, to derive the orbital parameters and the component masses.
Table \ref{table1.tab} lists the best-fit results.

To assess the ability of the GDR2/3 framework to recover orbital inclination from \textit{Gaia} absolute astrometry, we performed a comparative analysis of our derived results and the {\it Gaia} non-single-star (NSS) orbital solutions using a sample of systems from the {\it Gaia} NTBO catalog (See Supplementary Materials).
Overall, the two methods show good agreement (Figure \ref{compare_NSS_G23.fig}). The residual distributions for both inclination and $a_0$ are centered near zero, with mean offsets of only $\sim0.15^\circ$ ($\sigma=10.09^\circ$) and $\sim0.09$ mas ($\sigma=0.25$\,mas), respectively. These results indicate no significant systematic bias between the GDR2/3 method and the \textit{Gaia} NSS solutions.

Simultaneously, to account for potential unmodelled systematics (e.g., residual noise and possible frame-rotation effects) in J0008+0128, we added the standard deviation of the inclination residuals ($\approx$10$^\circ$) in quadrature to the fitting uncertainty ($\approx$5$^\circ$), yielding a final inclination angle of about $73^\circ\pm11^\circ$. 
Consequently, the mass of companion was derived as $2.04^{+0.30}_{-0.22}\ M_{\odot}$, $2.15^{+0.31}_{-0.22}\ M_{\odot}$, $2.35^{+0.35}_{-0.28}\ M_{\odot}$, $2.43^{+0.40}_{-0.34}\ M_{\odot}$, and $2.18^{+0.36}_{-0.30}\ M_{\odot}$, corresponding to different spectroscopic mass measurements of the visible star.
Based on these masses, we derived a best-fit mass of $2.20^{+0.14}_{-0.13}\ M_{\odot}$ for the compaion using Bayesian estimation (See Supplementary Materials).
Therefore, if the companion is indeed a compact object, it would likely be a massive or supermassive NS.
Figure \ref{fig2.fig} shows the best-fit achieved by the RV and astrometry models when applied to the data.

\begin{table*}
\centering
\small
\caption{{\bf Orbital parameters for J0008+0128 from the joint fitting to RV data and {\it Gaia} DR2 and DR3 data.} $M_{\rm 2}$ represents the updated companion mass after accounting for an inclination error of about 11$^\circ$. \label{table1.tab}}
\setlength{\tabcolsep}{1.2pt}
\renewcommand{\arraystretch}{0.95}
\begin{tabular}{p{3.5cm}lccccc}
\hline\noalign{\smallskip}
Parameter & Symbol & APOGEE & APOGEE & LAMOST & LAMOST & LAMOST \\
 &  & DR17 & SDSS-V & LRS & LRS & MRS \\
 &  & (ASPCAP) & (ASPCAP) & (LASP) & (DD-Payne) & (LASP) \\
\hline
\multicolumn{6}{l}{\bf{Parameters of the orbit}}  \\ 
Orbital period & $P$ [day] & $1218.71^{+3.11}_{-3.15}$ & $1218.62^{+3.24}_{-3.15}$ & $1218.60^{+3.20}_{-3.12}$ & $1218.57^{+3.25}_{-3.13}$ & $1218.53^{+3.24}_{-3.02}$ \\
RV semi-amplitude & $K_{1}$ [km/s] & $16.74^{+0.16}_{-0.16}$ & $16.74^{+0.16}_{-0.17}$ & $16.74^{+0.17}_{-0.16}$ & $16.74^{+0.16}_{-0.17}$ & $16.74^{+0.16}_{-0.17}$ \\
Systematic RV & $\nu{\rm 0}$ [km/s] & $-44.80^{+0.10}_{-0.10}$ & $-44.79^{+0.10}_{-0.10}$ & $-44.80^{+0.10}_{-0.10}$ & $-44.80^{+0.10}_{-0.10}$ & $-44.80^{+0.10}_{-0.10}$ \\
Eccentricity & $e$ [-] & $0.09^{+0.01}_{-0.01}$ & $ 0.09^{+0.01}_{-0.01}$ & $0.09^{+0.01}_{-0.01}$ & $0.09^{+0.01}_{-0.01}$ & $0.09^{+0.01}_{-0.01}$ \\
Inclination & $i$ [deg] & $73.17^{+5.59}_{-4.91}$ & $73.16^{+5.43}_{-5.14}$ & $73.04^{+5.38}_{-5.22}$ & $73.18^{+5.42}_{-4.99}$ & $73.12^{+5.56}_{-5.40}$ \\
Argument of periastron & $\omega$ [rad] & $0.69^{+0.11}_{-0.12}$ & $0.69^{+0.12}_{-0.12}$ & $0.69^{+0.11}_{-0.12}$ & $0.69^{+0.12}_{-0.12}$ & $0.69^{+0.12}_{-0.12}$ \\
Longitude of ascending node & $\Omega$ [rad] & $1.21^{+0.05}_{-0.05}$ & $1.21^{+0.05}_{-0.05}$ & $1.21^{+0.06}_{-0.05}$ & $1.21^{+0.06}_{-0.05}$ & $1.21^{+0.06}_{-0.05}$ \\
Mean anomaly at BJD 2456644.62918 & $M_{0}$ [rad] & $5.66^{+0.10}_{-0.09}$ & $5.67^{+0.10}_{-0.10}$ & $5.67^{+0.10}_{-0.09}$ & $5.67^{+0.10}_{-0.10}$ & $5.67^{+0.10}_{-0.10}$ \\
Re-normalized unit weight error & RUWE [-] & $1.615^{+0.190}_{-0.178}$ & $1.615^{+0.192}_{-0.175}$ & $1.622^{+0.197}_{-0.181}$ & $1.613^{+0.189}_{-0.175}$ & $1.616^{+0.198}_{-0.181}$ \\
\hline
\multicolumn{6}{l}{\bf{Astrometric offsets}}  \\ 
$\alpha^{*}$ offset & $\Delta \alpha^{*}$ [mas] & $-0.74^{+0.06}_{-0.06}$ & $-0.75^{+0.06}_{-0.06}$ & $-0.74^{+0.06}_{-0.06}$ & $-0.74^{+0.06}_{-0.06}$ & $-0.75^{+0.06}_{-0.06}$ \\
$\delta$ offset & $\Delta \delta$ [mas] & $-0.23^{+0.03}_{-0.03}$ & $-0.23^{+0.03}_{-0.03}$ & $-0.24^{+0.03}_{-0.03}$ & $-0.23^{+0.03}_{-0.03}$ & $-0.24^{+0.03}_{-0.03}$ \\
$\bar{\omega}$ offset & $\Delta \bar{\omega}$ [mas] & $0.00^{+0.03}_{-0.03}$ & $0.00^{+0.03}_{-0.03}$ & $0.00^{+0.03}_{-0.03}$ & $0.00^{+0.03}_{-0.03}$ & $0.00^{+0.03}_{-0.03}$ \\
$\mu_{\alpha^{*}}$ offset & $\Delta \mu_{\alpha^{*}}$ [mas/yr] & $1.31^{+0.05}_{-0.05}$ & $1.31^{+0.05}_{-0.05}$ & $1.31^{+0.05}_{-0.05}$ & $1.31^{+0.05}_{-0.05}$ & $1.31^{+0.05}_{-0.05}$ \\
$\mu_{\delta}$ offset & $\Delta \mu_{\delta}$ [mas/yr] & $0.84^{+0.11}_{-0.12}$ & $0.84^{+0.12}_{-0.12}$ & $0.84^{+0.13}_{-0.12}$ & $0.84^{+0.12}_{-0.12}$ & $0.84^{+0.12}_{-0.12}$ \\
\hline
\multicolumn{6}{l}{\bf{Jitter and error inflation}}  \\ 
RV jitter & $J_{\rm rv}$ [km/s] & $0.47^{+0.07}_{-0.06}$ & $0.47^{+0.07}_{-0.06}$ & $0.47^{+0.07}_{-0.06}$ & $0.47^{+0.07}_{-0.06}$ & $0.47^{+0.07}_{-0.06}$ \\
Error inflation of {\it Gaia} & $S$ [-] & $1.06^{+0.07}_{-0.04}$ & $1.06^{+0.07}_{-0.04}$ & $1.06^{+0.07}_{-0.04}$ & $1.06^{+0.07}_{-0.04}$ & $1.06^{+0.07}_{-0.04}$ \\
\hline
Mass of companion & $M_{\rm 2}$ [$M_{\odot}$] & $2.04^{+0.30}_{-0.22}$ & $2.15^{+0.31}_{-0.22}$ & $2.35^{+0.35}_{-0.28}$ & $2.43^{+0.40}_{-0.34}$ & $2.18^{+0.36}_{-0.30}$ \\
\hline
\end{tabular}
\end{table*}

\subsubsection*{Whether the visible component is a stripped star?}  
\label{sec:stripped_star}

The mass derived from the single-star models may be overestimated if the system undergoes a mass transfer.
Therefore, we investigated the possibility that the visible star of J0008+0128 is a stripped star in several aspects. 

(1) Is J0008+0128 currently undergoing mass transfer?

First, we calculated the size of Roche lobe for the visible star using the standard Roche lobe approximation \citep{1983ApJ...268..368E}:
\begin{equation}
\frac{R_{\rm roche}}{a} =  
    \frac{0.49 q^{2/3}}{0.6q^{2/3} + \ln(1 + q^{1/3})},
\label{eq7} 
\end{equation} 
where $a = (1+q) a_1 = (1+q) P (1-e^2)^{1/2} K_1/(2 \pi {\rm \sin}i)$ is the separation of binary system.
Taking the results from Bayesian estimation ($M_{\rm 1}=1.67^{+0.15}_{-0.15}\ M_{\odot}$ and $M_{\rm 2}=2.20^{+0.14}_{-0.13}\ M_{\odot}$), the filling factor of the visible star is $\approx $2$\%$, suggesting the absence of current mass transfer for J0008+0128.

Second, some binaries \citep{2022MNRAS.516.5945J,2022MNRAS.512.5620E} including a stripped star exhibit $H_{\alpha}$ emission lines, implying the presence of an accretion disk.
We checked the potential presence of a disk using the observed $H_{\alpha}$ line for J0008+0128.
Figure \ref{Ha.fig} shows the $H_{\alpha}$ profiles of J0008+0128 from LRS and MRS spectra in different phases.
No obviously emission profiles can be identified, suggesting the absence or weakness of accretion.

(2) Is J0008+0128 a post-mass transfer system?

First, some binaries \citep{2021MNRAS.502.3436E,2020A&A...639L...6S} containing a stripped star in its post-mass-transfer stage
exhibit $H_{\alpha}$ emission profiles. This emission line originates from a decretion disk surrounding the fast-rotating accretor.

Second, based on the current orbit, we estimated the masses of the visible star and the companion under the assumption that J0008+0128 is a post mass transfer binary.
Based on the orbital period and mass ratio derived from the Bayesian estimation, the Roche lobe of the visible star is estimated to be $\approx$260--280 $R_{\odot}$.
This excludes a low-mass stripped giant ($M \leq 0.6 M_{\odot}$), because it would not have expanded to such a large radius to overflow its Roche lobe \citep{1995MNRAS.273..731R}.
For a visible star with a mass of 0.6 $M_{\odot}$ and a inclination of 73$^{\circ}$, the inferred mass of the companion is about 1.38 $M_{\odot}$.  
Assuming the companion is a normal star, the flux ratios relative to the visible star in the in the $BP$, $G$, $RP$ and $H$ bands are 0.39, 0.31, 0.23 and 0.12, respectively. 
This means the visible companion can be detected in the SED or spectra.

We also examined the case in which the visible star is partially stripped. 
For example, based on the orbital solutions (Table \ref{keplerorbit.tab}), we computed the companion mass across a range of inclinations (e.g., $i=90^{\circ}$, $80^{\circ}$, and $70^{\circ}$), assuming residual masses of the visible star of $0.8$, $1.0$, and $1.2\,M_{\odot}$, respectively.
If the companion is a normal star, Table \ref{light_ratios.tab} lists the predicted flux ratios between the companion and the visible star in the {\it Gaia} ($G$, $BP$, $RP$) and 2MASS ($J$, $H$, $K_{s}$) bands.  
All predicted flux ratios indicate that such a companion would be detectable in the observed SED and spectra.

Third, for a post-mass transfer system, the effective temperature of stripped star would increase due to the shrinkage of the stripped star.
As an example, the temperature of the stripped star would increase by a factor of 2 to 3 when its radius shrinks to less than 0.1 of its Roche lobe.
In the case of J0008+0128, the ratio between the radius of visible star and its Roche lobe is about 2\%, suggesting the stripped star (post-mass transfer) would have an effective temperature greater than 12000 K \citep{2017MNRAS.467.1874C,2018A&A...615A..78G}.

As a result, the visible star of J0008+0128 is unlikely to be a (partially) stripped star, suggesting that the mass estimation is reasonable.

\subsubsection*{The nature of the dark component} 
\label{sec:nature_of_companion}

We investigated the possibility that the companion star of J0008+0128 is a normal star.
For the visible star, the total luminosity is about 16 $L_{\odot}$, while the expected total luminosity of a main-sequence star with a mass of 1.8--2.8 $M_{\odot}$ is between 11 $L_{\odot}$ and 81 $L_{\odot}$.
Therefore, if the companion star is a normal star, it would dominate or at least be a strong component of the spectrum and SED of J0008+0128.
However, the observed APOGEE spectra agree well with the best-fit spectrum from aspcapStar model, with no additional signatures detected (Figure \ref{fig1.fig}B).
On the other hand, a notable flux excess would be observed in optical bands if the companion were a normal star with a mass of 2.2 $M_{\odot}$ (Figure \ref{fig1.fig}C).
Consequently, the scenario in which J0008+0128 consists of two visible stars can be ruled out.

We also considered the possibility that the companion is a binary system.
For a Sun-like star, the absolute $U$, $B$, and $BP$ band magnitudes are 5.13 mag, 5.07 mag, and 4.67 mag.
If the companion consists of a Sun-like star and a low-mass compact object, the flux ratios between the companion and the visible star are 0.65, 0.15, and 0.11 in the $U$, $B$, and $BP$ bands, respectively, where the magnitudes of the visible star are 4.66, 2.98, and 2.30 mag in the corresponding bands.
If the companion instead consists of two Sun-like stars, the flux ratios rise to 1.30, 0.29, and 0.23 in the $U$, $B$, and $BP$ bands, respectively.
In both scenarios, a substantial excess should be apparent in the SED and optical spectra, contradicting the observed results (Figure \ref{fig1.fig}C).

If the companion consists of two low-mass compact objects --- such as two WDs, each with a mass of 1.1 $M_{\odot}$ --- we constrained the effective temperature of the WDs using the UV magnitudes (NUV and FUV) and WD cooling models (\url{https://www.astro.umontreal.ca/~bergeron/CoolingModels/}).
For the NUV magnitude, we adopted the reported value of ${\rm NUV}=20.039$ mag as documented by \citep{2011Ap&SS.335..161B}.
For the FUV band, we estimated an upper limit magnitude of $\approx$25.1–27.5 mag from {\it GALEX} images (See above data analysis).
A comparison between WD models \citep{2010MmSAI..81..921K} and the observed UV magnitude indicates that WDs with effective temperatures above 12,000 K can be excluded (Figure \ref{fig1.fig}C).
On the other hand, based on WD cooling models, a 1.1 $M_{\odot}$ WD requires approximately 1.7, 2.0, and 2.7 Gyr to cool to temperatures of 12,000, 11,000, and 10,000 K, respectively. However, the visible giant in J0008+0128 is estimated to be $2.68^{+0.90}_{-0.54}$ Gyr old (using APOGEE/DR17 parameters) or $2.40^{+0.64}_{-0.42}$ Gyr old (using APOGEE/SDSS-V parameters). This implies that a binary WD companion in this system must have a cooling temperature roughly in the range of 10,000 to 12,000 K to be consistent with the stellar age.
Such a narrow range of the temperature implies that the unseen object is unlikely to consist of two WDs.

Furthermore, evolutionary analysis suggests a low probability for a double WD scenario. On one hand, population synthesis of triple systems with inner WD+WD binaries indicates that the most compact configurations have outer orbits of dozens of AU \citep{2023ApJ...955L..14S}, much wider than J0008+0128 ($\sim$ 730 $R_{\odot}$ $\approx$ 3.4 AU). On the other hand, observations have not yet identified close double WDs with a total mass exceeding 1.4 $M_{\odot}$ \citep{2020A&A...638A.131N}, substantially lower than J0008+0128.

Based on the equation of state and some confirmed NS systems, several studies \citep{2017ApJ...850L..19M,2018ApJ...852L..25R,2019PhRvD.100b3015S,2020MNRAS.499L..82M,2025NatAs...9..552Y} argued that the maximum mass of an NS is between 2.15 $M_{\odot}$ and 2.3 $M_{\odot}$,
which falls within the mass range of the companion of J0008+0128. 
Consequently, if the companion of J0008+0128 is indeed a compact object, it is most likely a supermassive NS (or a low-mass BH).

\subsubsection*{Spectral disentangling} 
\label{sec:spectral_disentangling}

Spectral disentangling is a reliable method to detect the additional component in spectra.
We performed the spectral disentangling using the algorithm proposed by Simon and Sturm \citep{1994A&A...281..286S} to the APOGEE spectra.
Assuming the component spectra remain constant over time, the algorithm is designed to solve for linear functions constructed from the spectral data.
Assuming the companion is a normal star with a mass of 2.2 $M_{\odot}$, the flux ratio between the companion and the visible star in the $H$ band is about 0.38, suggesting the companion is detectable through spectral disentangling.
We used the wavelength range from 16660 \AA \ to 16800 \AA \ for the spectral disentangling process.

Following the method of \citep{2024NatAs...8.1583W,2024ApJ...977..245Z}, we first investigated the detection limit of the companion for J0008+0128.
For the synthesis of binary spectra, we used mass ratios of $q =$ 0.6, 0.65, 0.70, 0.75 and 0.8.
The rotational broadening ($v{\rm sin}i$) of the companion was set at 50, 100, and 150 km/s, respectively.
Consequently, we found that all systems can be disentangled well, indicating that the spectral disentangling algorithm is feasible for J0008+0128.

Considering the mass ratios of 0.72, 0.77, and 0.79 (corresponding to the inclinations of 70$^{\circ}$, 80$^{\circ}$, and 90$^{\circ}$), we disentangled the observed APOGEE spectra of 0008+0128.
Similar to the method described by Seeburger et al. \citep{2024MNRAS.530.1935S}, we estimated the light ratio $\alpha$ using the reduced chi-square $\tilde{\chi}^2$ between the templates and the component spectra as follows:
\begin{equation}
\tilde{\chi}^2 = \sum\frac{(\frac{\alpha}{1+\alpha}t^{\rm A} - x^{\rm A} + \frac{1}{1+\alpha}t^{\rm B} - x^{\rm B})^{2}}{\sigma^{2}},
\label{light_ratio} 
\end{equation} 
where the $t^{\rm A}$ and the $t^{\rm B}$ are the templates for both components, while the $x^{\rm A}$ and the $x^{\rm B}$ represent the component spectra. $\sigma$ is the error of flux.

We used Equation \ref{light_ratio} to estimate the light ratio $\alpha$ between the two component spectra obtained from spectral disentangling.
Assuming the companion is a main-sequence star, we derived its atmospheric parameters (i.e., $T_{\rm eff}$ and log$g$) (\url{http://www.pas.rochester.edu/~emamajek/EEM_dwarf_UBVIJHK_colors_Teff.txt}).
Then, we used the Phoenix model with similar atmospheric parameters as the template (i.e., $t^{\rm A}$ and $t^{\rm B}$) for each component.
For the light ratio $\alpha$, we adopted a grid ranging from 0 to 100 with a step size of 0.01 to calculate the corresponding $\tilde{\chi}^2$.
As an example, Figure \ref{luminosity_ratio.fig} plots the light ratio $\alpha$ and the corresponding $\tilde{\chi}^2$.
The $\tilde{\chi}^2$ value decreases as the light ratio $\alpha$ increases, indicating the contribution of the second component spectrum to the observed spectra is negligible.
Finally, no obvious additional component can be found for those mass ratios.
Figure \ref{disentangling_results.fig} plots the disentangling results using the mass ratio of $q=$0.79.
The results obtained from the spectral disentangling further support the hypothesis that the companion of J0008+0128 is a massive compact object.


\clearpage 

\bibliography{refs}
\bibliographystyle{sn-nature}

\newpage


\section*{Acknowledgments}

We thank the referees for a careful review of the manuscript and providing constructive feedback for improving its contents. 
We thank Dr. Zikun Lin, Heyang Liu, Huaqing Cheng for very useful discussions.
This work made use of the data from LAMOST (Large Sky Area Multi-Object Fiber Spectroscopic Telescope, also known as the Guoshoujing Telescope) (https://cstr.cn/31118.02.LAMOST). LAMOST is a Chinese national mega-science facility, operated by National Astronomical Observatories, Chinese Academy of Sciences.
Funding for the Sloan Digital Sky Survey (III, IV, and V) has been provided by the Alfred P. Sloan Foundation, the Heising-Simons Foundation, the National Science Foundation, and the Participating Institutions. SDSS acknowledges support and resources from the Center for High-Performance Computing at the University of Utah. SDSS telescopes are located at Apache Point Observatory, funded by the Astrophysical Research Consortium and operated by New Mexico State University, and at Las Campanas Observatory, operated by the Carnegie Institution for Science. 
This work presents results from the European Space Agency (ESA) space mission {\it Gaia}. {\it Gaia} data are being processed by the {\it Gaia} Data Processing and Analysis Consortium (DPAC). Funding for the DPAC is provided by national institutions, in particular the institutions participating in the {\it Gaia} MultiLateral Agreement (MLA). 
This work made use of the data from FAST (Five-hundred-meter Aperture Spherical radio Telescope). FAST is a Chinese national mega-science facility, operated by National Astronomical Observatories, Chinese Academy of Sciences.
This work made use of the data from Einstein Probe, a space mission supported by Strategic Priority Program on Space Science of Chinese Academy of Sciences, in collaboration with ESA, MPE and CNES. 

\paragraph*{Funding:}
This work was funded by the Strategic Priority Program of the Chinese Academy of Sciences (SPPCAS) under grant number XDB1160302 (S.W.), the National Natural Science Foundation of China (NSFC) under grant number 12588202 (J.L.), the National Key Research and Development Program of China under grant number 2023YFA1607901 (S.W.), Chinese Academy of Sciences Basic Frontier Science Research Program from 0 to 1 Original Innovation Project under grant No. ZDBS-LY-SLH002 (J.L.), SPPCAS under grant number XDB1160101 (H.Y.),  the NSFC under grant number 12273057 (S.W.), science research grants from the China Manned Space Project (S.W.), the New Cornerstone Science Foundation through the New Cornerstone Investigator Program and the XPLORER PRIZE (J.L.), Chongqing Natural Science Foundation under grant number CSTB2023NSCQ-MSX0093 (X.Y.), the NSFC under grand number 12547101 (X.Z.), the Postdoctoral Fellowship Program of China Postdoctoral Science (CPSF) under grant number GZB20250739 (X.Z.), and the China Postdoctoral Science Foundation under grant number 2026M793519 (X.Z.).

\paragraph*{Author contributions:}
Conceptualization: S.W., J.L., X.Z.\\
Methodology: X.Z., G.X., M.S., K.R., S.G., H.Y., M.Z., F.F.\\
Software: X.Z., M.Z.\\
Validation: X.Z., S.W., R.S., G.X., M.S., S.G., J.L., F.F., Y.H.\\
Formal analysis: X.Z., G.X., S.W., M.S., K.R., S.G., X.M., P.W., H.Y., M.Z., X.Li., S.L.\\
Investigation: S.W., J.L., X.Z., G.X., M.S., K.R., S.G., X.M., P.W., M.Z., X.Luo, M.X., Y.H., F.F., S.L.\\
Resources: S.W., J.L., D.Q.\\
Data curation: X.Z., D.Q., S.L.\\
Writing – original draft: X.Z., S.W., P.W.\\
Writing – review \& editing: X.Z., S.W., R.S., G.X., M.S., S.G., X.M., X.Li., X.Y., J.L., F.F., Y.H.\\
Visualization: X.Z.\\
Supervision: S.W., J.L.\\
Project administration: S.W., J.L.\\
Funding acquisition: S.W., J.L.
\paragraph*{Competing interests:}
The authors declare they have no competing interest.
\paragraph*{Data, Code, and Materials Availability:} 
This study did not generate new materials.
All data and code needed to evaluate and reproduce the results in the paper are present in the paper and/or the Supplementary Materials. 
Spectroscopic data from LAMOST DR11 are available for download at \url{http://lamost.org/dr11/}.
Spectroscopic data from APOGEE DR17 and SDSS-V are available at \url{https://data.sdss.org/sas/}.
The source code for spectral disentangling is available at Zenodo (\url{https://zenodo.org/records/22114414}) or Github (\url{https://github.com/Xinlin-code/Spectral-disentangling}), and the source code for the joint RV and astrometric fitting is provided at Zenodo (\url{https://zenodo.org/records/22114147}) or Github (\url{https://github.com/Xinlin-code/GDR2-3-method-for-J0008-0128}).

\paragraph{List of Supplementary Materials:}
Sections S1 to S9\\
Tables \ref{rvdata.tab} to \ref{other_ns.tab}\\
Figures \ref{compare_NOT_template.fig} to \ref{2n_3n.fig}
\clearpage 


\renewcommand{\thefigure}{S\arabic{figure}}
\renewcommand{\thetable}{S\arabic{table}}
\renewcommand{\theequation}{S\arabic{equation}}
\renewcommand{\thepage}{S\arabic{page}}
\setcounter{figure}{0}
\setcounter{table}{0}
\setcounter{equation}{0}
\setcounter{page}{1} 


\begin{center}
\section*{Supplementary Materials for\\ \scititle}
    Xinlin Zhao,
    Song Wang$^{\ast}$,
    Guang-Yao Xiao,
    Meng Sun,
    Roberto Soria,
    Kyle Akira Rocha,
    Seth Goggage,
    Pengfei Wang,
    Hongliang Yan,
    Xuan Mao,
    Shuai Liu,
    Yang Huang,
    Maosheng Xiang,
    Dan Qiu,
    Meng Zhang,
    Xue Li,
    Fabo Feng,
    Xihong Luo,
    Xiaohong Yang,
    Jifeng Liu$^{\ast}$ \\
	\small$^\ast$Corresponding author. Email: songw@bao.ac.cn; jfliu@nao.cas.cn
\end{center}

\subsubsection*{This PDF file includes:}
Supplementary Sections S1 to S9\\
Supplementary Tables \ref{rvdata.tab} to \ref{other_ns.tab}\\
Supplementary Figures \ref{compare_NOT_template.fig} to \ref{2n_3n.fig}\\
References\\

\clearpage

\subsubsection*{S1. Spectroscopic observations and data reduction}
\label{sec:observation_data_reduction}

Combining the spectroscopic data from Large Sky Area Multi-Object Fiber Spectroscopic Telescope (LAMOST) and Apache Point Observatory Galactic Evolution Experiment (APOGEE), as well as the orbital solutions from {\it Gaia} DR3, we conducted a search for massive compact objects.
First, we cross-matched the NTBO table of {\it Gaia} DR3 with LAMOST and APOGEE data. 
Second, we calculated the mass function using the orbital solutions from the NTBO table.
Third, we selected the sources with multiple ($>$ 2) spectral observations and RV variation larger than 30 km/s.
We identified a wide binary {\it Gaia} ID 2546630699873483008 (R.A. = 2.00510$^\circ$; Decl. = 1.48284$^\circ$; hereafter J0008+0128), whose large binary mass function suggests the possible presence of a massive compact object.

APOGEE is a large-scale and stellar spectroscopic survey aimed at exploring galactic evolution.
The spectroscopic data were obtained by two high-resolution ($\Delta \lambda / \lambda \sim$ 22500 \AA) and multi-fibre spectrographs \citep{2019PASP..131e5001W}, which are mounted on the 2.5m Sloan telescope at Apache Point Observatory (APO) \citep{2006AJ....131.2332G} and the du Pont 2.5m telescope at Las Campanas Observatory (LCO) \citep{1973ApOpt..12.1430B}.
The spectra were reduced using a customized pipeline \citep{2015AJ....150..173N}, with a wavelength range of 15000 \AA \ to 17000 \AA.
The atmospheric parameters of each spectrum are estimated using the {\it ASPCAP} pipeline \citep{2016AJ....151..144G}.
From Dec. 2013 to Oct. 2022, J0008+0128 was observed seven times in APOGEE DR17 (three observations in SDSS-III and four observations in SDSS-IV) and nine times in APOGEE/SDSS-V.
All spectra obtained from APOGEE DR17 and SDSS-V were used for the following analyses.

LAMOST has begun its second five-year spectroscopic survey, using both low-resolution (LRS, $\Delta \lambda / \lambda \sim$ 1800 \AA) and medium-resolution (MRS, $\Delta \lambda / \lambda \sim$ 7500 \AA) observations. 
The LRS spectra cover a wavelength range from 3690 \AA \ to 9100 \AA \ \citep{2015RAA....15.1095L}, while the MRS spectra are divided into two arms: the blue arm spans wavelengths from 4950 \AA \ to 5350 \AA, and the red arm covers wavelengths from 6300 \AA \  to 6800 \AA \ \citep{2020arXiv200507210L}.
In LAMOST DR11, a total of 12 million LRS spectra and 13 million MRS spectra have been collected, establishing a comprehensive database for the identification of compact binaries.
We obtained the LRS and MRS spectra for J0008+0128, including 3 LRS and 6 MRS.
Among the LRS and MRS spectra, only the spectra with S/N greater than 10 are used for the following analyses.

Furthermore, we obtained two consecutive high‑resolution spectra on 18 December 2025 with the FIbre‑fed Echelle Spectrograph (FIES) on the Nordic Optical Telescope (NOT). The observations were taken at a resolving power of 
R$\approx$67,000 (\url{https://www.not.iac.es/instruments/fies/}).

We utilized the cross-correlation function (CCF) to estimate the RV values for each spectrum. 
The Phoenix (\url{ftp://phoenix.astro.physik.uni-goettingen.de/}) model was adopted and re-sampled to match the resolutions of the APOGEE, LAMOST, and NOT spectra for RV measurements. 
For the APOGEE spectra, we selected the observed spectra within the wavelength range of 15150 \AA \ to 15700 \AA. 
For the LRS spectra, wavelengths from 5000 \AA \ to 6000 \AA \ were used. 
In the case of MRS spectra, two wavelength ranges were employed: 4950 \AA \ to 5350 \AA \ (blue band) and 6350 \AA \ to 6850 \AA \ (red band).
For NOT spectra, the wavelength range from 5500 \AA\ to 6800 \AA\ was employed.

Furthermore, we determined the RV zero point (RVZP) for each APOGEE and LAMOST spectrum following Wang et al. \citep{2021RAA....21..292W}. 
For each spectrum, we selected single stars observed on the same day with the same spectrograph, derived the RV offset for each star by comparing their RVs (from CCF measurements) with the {\it Gaia} DR3 reference value, and took the median RV offset of several dozen stars as the final RVZP value.
Table \ref{rvdata.tab} lists the RVZP-corrected RV data, with the uncertainty calculated as the square root of the sum of the squares of the measurement error and the error of RVZP.

\subsubsection*{S2. Atmospheric parameters and their uncertainties}  
\label{sec:at_parameters}

The stellar parameters of J0008+0128 were first compiled from the APOGEE and LAMOST parameter catalogs. For each dataset, we adopted the (signal-to-noise-ratio-weighted) average of multiple observations as the initial parameter value and the standard deviation as its uncertainty.
As shown in Table \ref{atmo_pars.tab}, the effective temperature ($T_{\rm eff}$) and metallicity ([Fe/H]) are consistent across all datasets, while the surface gravity (log$g$) inferred from LAMOST spectra is slightly higher than that from APOGEE. 
Although we obtained optical high‑resolution spectra with NOT, the low signal‑to‑noise ratio (S/N) prevented a reliable determination of atmospheric parameters from Fe I and Fe II lines. Figure \ref{compare_NOT_template.fig} shows the co‑added NOT spectrum alongside a synthetic template computed using the APOGEE DR17 parameters.

To correct for systematic offsets between different spectroscopic surveys, we used reference datasets: the infrared flux method for $T_{\rm eff}$, astroseismology for log$g$, and open cluster measurements for [Fe/H].
For each survey, we selected stars with atmospheric parameters similar to the visible star of J0008+0128, and used the common stars found in the reference sample to compute the median offset and its robust scatter (half-width of the central 68\% interval). The derived offset is then added to the initial parameter value to remove the systematic bias.
Simultaneously, the initial uncertainty provided by each survey reflects only the random error ($\sigma_r$).
To account for systematic uncertainties in the spectroscopic measurements and obtain a more complete error budget, we incorporated the method error ($\sigma_m$), defined as the robust scatter from the comparison with the reference sample.

Such a method error may be subject to additional effects, including random errors from both catalogs, astrophysical contamination, and parameter-dependent biases. To mitigate the impact of random errors, we restricted our comparison to spectra with S/N higher than 50. We also cross-matched the common stars with SIMBAD to identify non-stellar objects (e.g., galaxies, AGNs) and binaries. The small number of such contaminants is unlikely to affect $\sigma_m$. Finally, by restricting our analysis to stars with similar atmospheric parameters, the narrow parameter ranges make parameter-dependent biases negligible.

We adopted the catalog from Zhang et al. \citep{2025ApJS..279....5Z} as our reference for $T_{\rm eff}$, as the $T_{\rm eff}$ in this catalog have been calibrated with the infrared flux method.
Stars with S/N$>50$, $4700$ K $<T_{\rm eff}< 5300$ K, $3.0<$log$g<3.5$, and $-0.5<$[Fe/H]$<0$ are selected from each dataset to calculate the $T_{\rm eff}$ offsets.
We obtained the median offsets of 9 K, 71 K, $-$16 K, and 6 K for APOGEE/DR17 (ASPCAP), APOGEE/SDSS-V (ASPCAP), LAMOST/LRS (LASP), and LAMOST/MRS (LASP), respectively. The corresponding $\sigma_m$ values are 39 K, 38 K, 32 K, and 59 K, respectively.

Accurate log$g$ derived from asteroseismology provide a gold-standard reference dataset for spectroscopic log$g$ estimates.
We selected stars with S/N$>50$, $4700$ K $<T_{\rm eff}< 5300$ K, $3.0<$log$g<3.5$, and $-0.5<$[Fe/H]$<0$ from each dataset to calculate the log$g$ offsets.
By combining the frequency of maximum oscillation power ($\nu_{\rm max}$) from Yu et al. \citep{2018ApJS..236...42Y} with the $T_{\rm eff}$ from various datasets, we estimated the asteroseismic log$g$ values as follows \citep{1995A&A...293...87K}:
\begin{equation}
    {\rm log}g = {\rm log}g_{\odot} + {\rm log}(\frac{\nu_{\rm max}}{\nu_{\rm max, \odot}}) + \frac{1}{2} {\rm log}(\frac{T_{\rm eff}}{T_{\rm eff, \odot}}).
\label{mf}
\end{equation}
\noindent
For the solar frequency max ($\nu_{\rm max, \odot}$), a value of 3090 $\mu$Hz is adopted \citep{2011ApJ...743..143H}.
A comparison with these asteroseismic log$g$ values yielded the median offsets of $-$0.01, $-$0.01, $-$0.08, 0.01, and $-$0.06 for APOGEE/DR17 (ASPCAP), APOGEE/SDSS-V (ASPCAP), LAMOST/LRS (LASP), LAMOST/LRS (DD-Payne), and LAMOST/MRS (LASP), respectively. The derived $\sigma_m$ values are 0.06, 0.05, 0.07, 0.07, and 0.09, respectively.

Since the member stars in open clusters share nearly identical metallicities, we utilized several well-studied clusters to estimate the $\sigma_m$ for [Fe/H] across the different datasets. 
We collected clusters with metallicities spanning $-0.5$ to $0.1$, including NGC 188, NGC 1245, NGC 2204, NGC 2158, NGC 2420, NGC 7789, NGC 6819, and NGC 2682, from \citep{2020AJ....159..199D}. To ensure a sufficiently large common star sample, we slightly relaxed the selection criteria, requiring S/N$>30$, $4500$ K $<T_{\rm eff}< 6000$ K, and $3.0<$log$g<4.0$.
We obtained the median offsets of 0.01, 0.01, 0.01, 0.07, and 0.02 for APOGEE/DR17 (ASPCAP), APOGEE/SDSS-V (ASPCAP), LAMOST/LRS (LASP), LAMOST/LRS (DD-Payne), and LAMOST/MRS (LASP), respectively.
The corresponding $\sigma_m$ values are 0.05, 0.06, 0.06, 0.06, and 0.05, respectively.

Finally, we adopted $\sqrt{\sigma_{\rm r}^{2} + \sigma_{\rm m}^{2}}$ as the final uncertainties for the atmospheric parameters. 
Table \ref{atmo_pars.tab} summarizes the corrected results for J0008+0128.
Our estimated errors are consistent with previous studies. For APOGEE, \citep{2016ApJ...823..114N} reported errors of 31 K, 0.07 dex, and 0.02 dex for $T_{\rm eff}$, log$g$, and [Fe/H], respectively, while \citep{2025ApJS..276...69P} provided errors of 49 K, 0.07 dex, and 0.05 dex.
For LAMOST, \citep{2022ApJS..259...51W} reported errors of 85 K, 0.1 dex, and 0.05 dex for $T_{\rm eff}$, log$g$, and [Fe/H], respectively, while \citep{2025ApJS..279....5Z} found 52 K, 0.11 dex, and 0.06 dex, respectively.

\subsubsection*{S3. Bayesian mass estimates for the visible star and companion}
\label{sec:mass_estimate}

We performed a Bayesian analysis to derive the mass likelihood $P({\rm data}|M)$ for both the visible star and its companion using the five measurements reported in Table \ref{table1.tab}. 
We used {\it PyMC} \citep{pymc2023}, a Python package for probabilistic programming and Bayesian analysis, to build a single-parameter Bayesian model. 
A uniform prior in the range [1, 3] $M_{\odot}$ was adopted for both components. 
The posterior distribution was sampled via MCMC using the No-U-Turn Sampler algorithm, with 5000 sampling iterations and 1000 tuning iterations. 
The 50th, 16th, and 84th percentiles of the posterior distribution were adopted as the best-fit mass and its uncertainties. As a result, we obtained a best estimate of $1.67^{+0.15}_{-0.15}\ M_{\odot}$ for the visible star and $2.20^{+0.14}_{-0.13}\ M_{\odot}$ for the companion. 
The corresponding mass estimates are shown in Figure~\ref{Bayesian_estimate_for_mass.fig}.

\subsubsection*{S4. Estimate of the renormalised unit weight error}
\label{sec:ruwe}

J0008+0128 has a renormalised unit weight error (RUWE) of 1.309 from {\it Gaia} DR3 catalog, slightly below the widely used threshold of 1.4 that typically indicates an unresolved binary.
This can be attributed to the companion's orbital period exceeding the {\it Gaia} baseline, or to an underestimation of the astrometric uncertainties.
To explore whether our orbital solution can explain the reported RUWE, we simulated the RUWE value following these formula (\url{http://www.rssd.esa.int/doc_fetch.php?id=3757412}): 
\begin{eqnarray}
{\rm RUWE}=\frac{{\rm UWE}}{u_{0}},
\label{ruwe}
\end{eqnarray}  
\begin{eqnarray}
{\rm UWE}=\sqrt{\frac{\chi^2_{\rm unbinned}}{\nu}},
\label{uwe}
\end{eqnarray}
and
\begin{eqnarray}
\chi^2_{\rm unbinned}=\chi^2_{\rm binned}+N_{\rm FOV\,transits}\,(N_{\rm bin}-1),
\end{eqnarray}
where UWE is the unit weight error and $u_{0}$ is the re-normalized factor. The $\chi^2_{\rm unbinned}$ is the value calculated for all CCD along-scan data points, $\nu$ ($=N_{\rm FOV\,transits}\,N_{\rm bin}-5$) is the number of degrees of freedom. The GOST predictions can be treated as the binned data, and it's $\chi^2_{\rm binned}$ can be converted to unbinned $\chi^2_{\rm unbinned}$. 
Assuming $N_{\rm bin}=8$, the UWE values for the solutions derived from APOGEE/DR17 (ASPCAP), APOGEE/SDSS-V (ASPCAP), LAMOST/LRS (LASP), LAMOST/LRS (DD-Payne), and LAMOST/MRS (LASP) are 1.972, 1.973, 1.981, 1.969, and 1.974, respectively. Taking the number of AL observations after outlier rejection (\texttt{astrometric\_n\_good\_obs\_al}) and the $\chi^2$ (\texttt{astrometric\_chi2\_al}) along with the RUWE from {\it Gaia} DR3, we computed a renormalised factor of $u_{0}=1.221$. Applying this factor, the final RUWE values are 1.615, 1.615, 1.622, 1.613, and 1.616, respectively (listed in Table \ref{table1.tab}).

Our simulated RUWE is roughly comparable with the reported one, suggesting our orbital solution for the dark companion is reasonable. The small discrepancy is likely due to the mismatch between the number of simulated and actual observations, as well as to any unmodelled astrometric uncertainties.
Moreover, the uncertainty in the {\it Gaia} along-scan abscissa measurements \citep{2023A&A...674A..10H} can significantly influence the estimated RUWE values. For example, introducing a calibration noise of only 0.04 mas in quadrature to the field-of-view (FOV) transit uncertainties can reduce the predicted RUWE to approximately 1.3, which aligns well with the observed value.

Additionally, we obtained a predicted value of ${\rm RUWE}=2.16$ using {\it gaiamock} \citep{2024OJAp....7E.100E}, which significantly deviates from both our estimated value and the observed value.
This difference can be attributed to several main factors.

First, the output of {\it gaiamock} is a ``scalable" UWE. To enable a direct comparison with observed RUWE, {\it gaiamock} adds a Gaussian noise with a mean of zero and standard deviation of one, resulting in a Gaussian distribution centered on the UWE. Moreover, for systems with $G_{\rm mag} < 13$ mag, {\it gaiamock} introduces an unmodeled Gaussian noise term, $\mathcal{N}(0, 0.04)$, to account for underestimated uncertainties due to various systematics. They found that adding 0.04 mas in quadrature to the per-FOV transit uncertainties yields a reduced $\chi^2$ of $\approx$1 for {\it Gaia} BH3 \citep{2024OJAp....7E.100E}. However, these introduce two problems: the scalable UWE given by {\it gaiamock} isn't equivalent to RUWE; the assumed 0.04 mas value may not be appropriate for all systems, potentially introducing additional bias or uncertainty into the RUWE estimates.
Interestingly, {\it gaiamock} works well to provide RUWE estimates for a wide range of systems \citep{2025OJAp....8E..62E}.
This is because that the re-normalized factor $u_{0}$ (in Eq. \ref{ruwe}) is approximately unity for the majority of systems (See the official documentation of the {\it Gaia} Data Processing and Analysis Consortium (\url{http://www.rssd.esa.int/doc_fetch.php?id=3757412})), which means the UWE can serve as a statistically reliable approximation to the RUWE. 
Furthermore, a comparison between the {\it gaiamock} prediction and the observed RUWE reveals a scatter of approximately 17\%, and a slight bias toward higher predicted than observed RUWE has also been noticed in the low-RUWE regime \citep{2025OJAp....8E..62E}.
Consequently, for individual systems, an accurate re-normalization factor is necessary to achieve a more reliable RUWE estimate.

Second, unlike {\it gaiamock} which rejects GOST predictions with 10\% probability to model {\it Gaia} abscissae, we directly took into account the publicly available gaps (\url{https://www.cosmos.esa.int/web/gaia/dr3-data-gaps}) to remove bad FOV transits and rescaled $\chi^2_{\rm binned}$. Therefore, we think that our result for individual sources is more accurate than {\it gaiamock}.

\begin{figure}
    \center
    \includegraphics[width=0.98\textwidth]{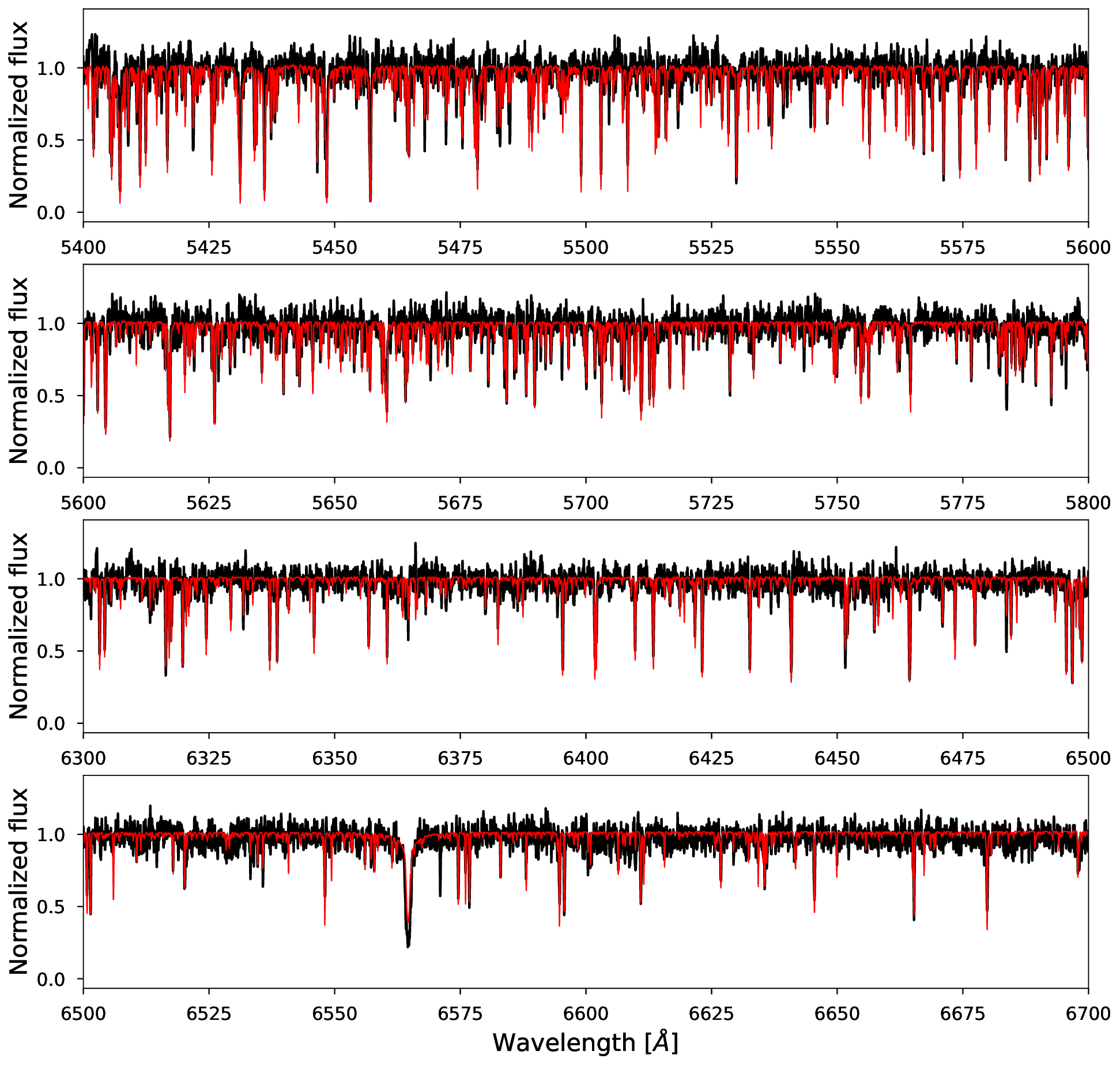}
    \caption{{\bf Comparison between the co-added NOT spectrum and the PHOENIX model.} The model was generated using stellar parameters derived from APOGEE DR17. The black curve represents the co-added NOT spectrum, while the red curve corresponds to the PHOENIX template.}
    \label{compare_NOT_template.fig}
\end{figure}

\begin{figure}
    \center
    \includegraphics[width=0.98\textwidth]{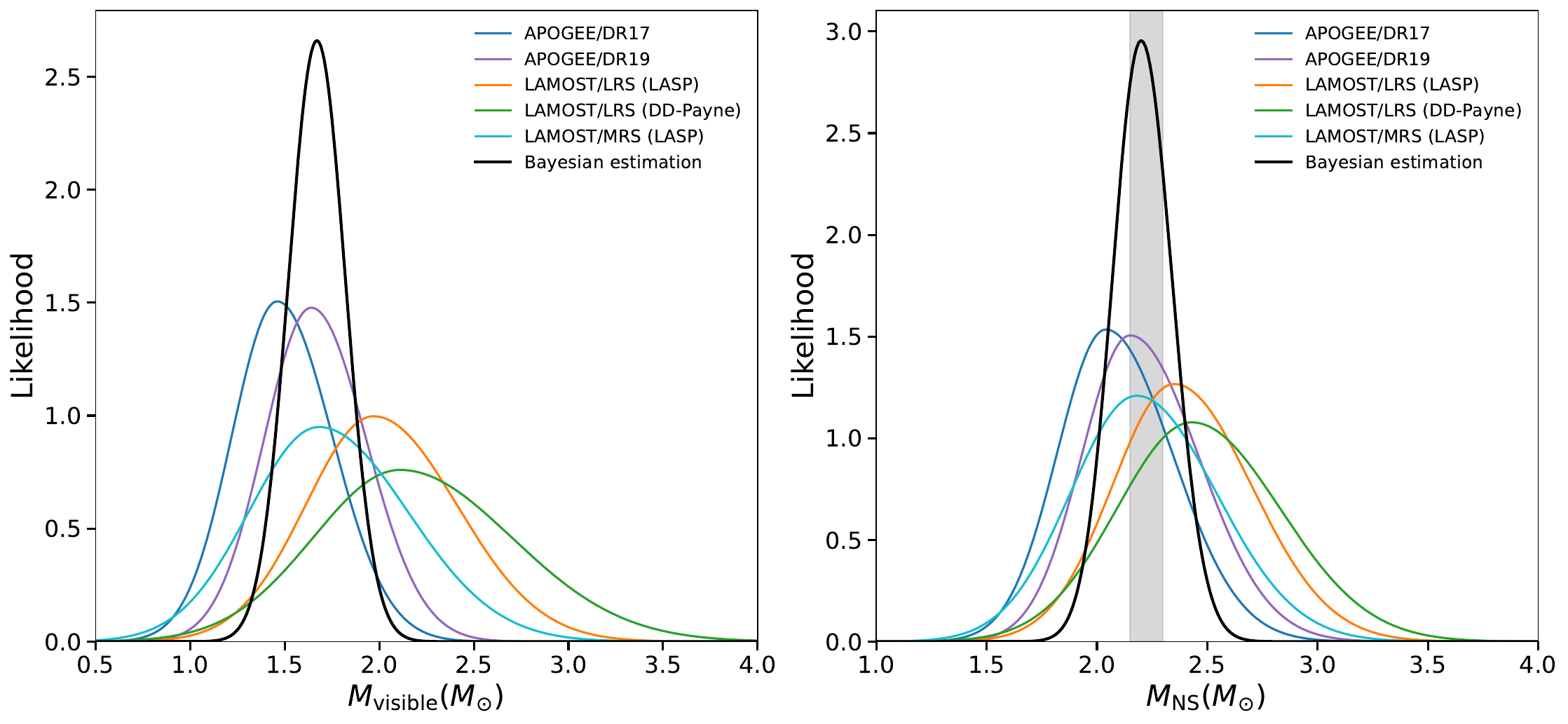}
    \caption{{\bf Bayesian mass estimates for the visible star (left) and the companion (right) from different methods.} The gray shaded region indicates the maximum mass range (2.15--2.30 $M_{\odot}$) for non-rotating or slowly rotating NSs based on the equations of state
    \citep{2017ApJ...850L..19M,2018ApJ...852L..25R,2019PhRvD.100b3015S}.}
    \label{Bayesian_estimate_for_mass.fig}
\end{figure}

In addition, {\it gaiamock} returns a flux ratio in the $G$-band of $\sim$0.5 for J0008+0128. In such a case, a hot companion should be readily detectable in optical band. However, the CCF between the NOT high-resolution spectra and the template show a clean single-peaked profile, ruling out the presence of a clearly visible companion. This conclusion is further supported by SED fitting (Figure \ref{fig1.fig}) and direct comparison between the co-added NOT spectrum and synthetic templates (Figure \ref{compare_NOT_template.fig}).

\subsubsection*{S5. Validation of Our Results against {\it Gaia} NSS Orbital Solutions}
\label{sec:lc}

\begin{figure}
    \center
    \includegraphics[width=0.98\textwidth]{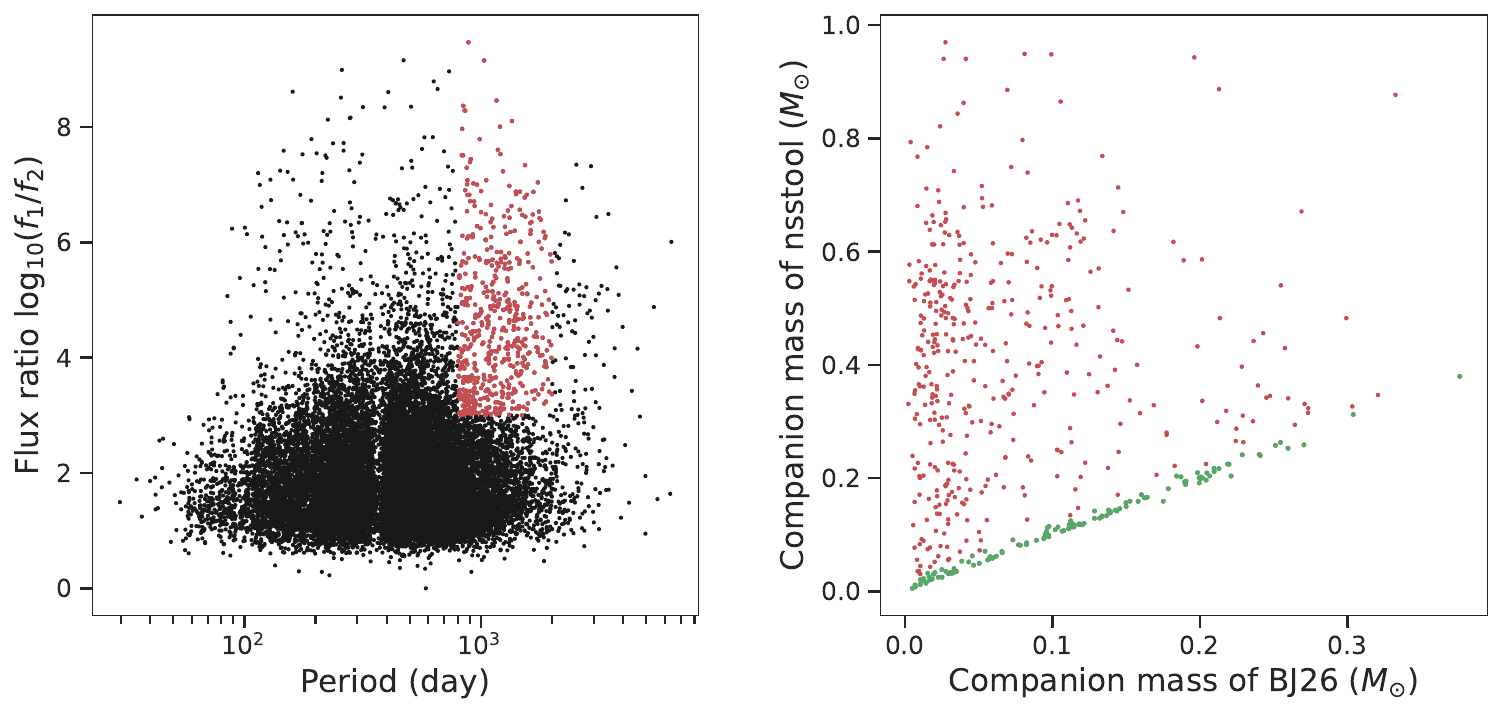}
    \caption{{\bf Sample selection for the comparison between our method and the {\it Gaia} NSS orbital solutions.} Left panel: Orbital period versus flux ratio between the brighter and fainter components in the $G$ band.
     Black dots represent the full sample from BJ26, while red dots denote targets that satisfy criteria (a) and (b). Right panel: Comparison of masses derived from nsstool and those reported in BJ26. Red points indicate systems meeting criteria (a) and (b), and green points correspond to systems that further satisfy criterion (c).
    \label{select_sample.fig}}
\end{figure}

To assess the capability of the GDR2/3 framework to recover orbital inclination from \textit{Gaia} absolute astrometry, we conducted a comparative analysis between our method and \textit{Gaia} NTBO results. Direct application of our method to all NTBO targets is impractical because the method is specifically designed for systems hosting dark companions. For binaries with luminous companions, the modelling of flux contribution from the companion may introduce additional degeneracies among orbital parameters, especially given the limited number of astrometric measurements available from GDR2/3. Instead, we constructed a validation sample based on \citep{BJ2026A&A} (hereafter BJ26), who derived individual component masses for unresolved binaries by combining NSS orbital solutions with three-band \textit{Gaia} photometry. They estimated the flux ratios using mass-flux relations calibrated from stellar isochrones in each \textit{Gaia} band. Our selection criteria are: (a) flux ratio between the brighter and fainter components $f_{\rm 1}/f_{\rm 2}>1000$; (b) orbital period $800<P<2000$ d; and (c) consistency between the companion masses derived by BJ26 and \texttt{nsstool} (\url{https://gitlab.obspm.fr/gaia/nsstools}) \citep{Halbwachs2022}, i.e., $|M_{\rm BJ26}-M_{\rm nss}|<0.02 M_{\odot}$ (Figure~\ref{select_sample.fig}). The \texttt{nsstool} converts the Thiele-Innes coefficients into the corresponding Campbell elements, from which the companion mass can be inferred once the mass of the brighter star is specified. Criterion (a) conservatively selects systems likely hosting dark companions, while criterion (b) broadly matches both the \textit{Gaia} temporal baseline and the sensitivity range of our method \citep{Xiao2026ApJ}. Criterion (c) further excludes systems with potentially luminous or ambiguous companions, since significant discrepancies between the BJ26 and \texttt{nsstool} masses may indicate unreliable mass estimates or non-negligible flux contamination from the companion, thereby violating the dark-companion assumption adopted in our method. As shown in the right panel of Figure~\ref{select_sample.fig}, several systems indeed exhibit discrepant companion masses.

\begin{figure}
    \centering
    \includegraphics[width=0.49\textwidth]{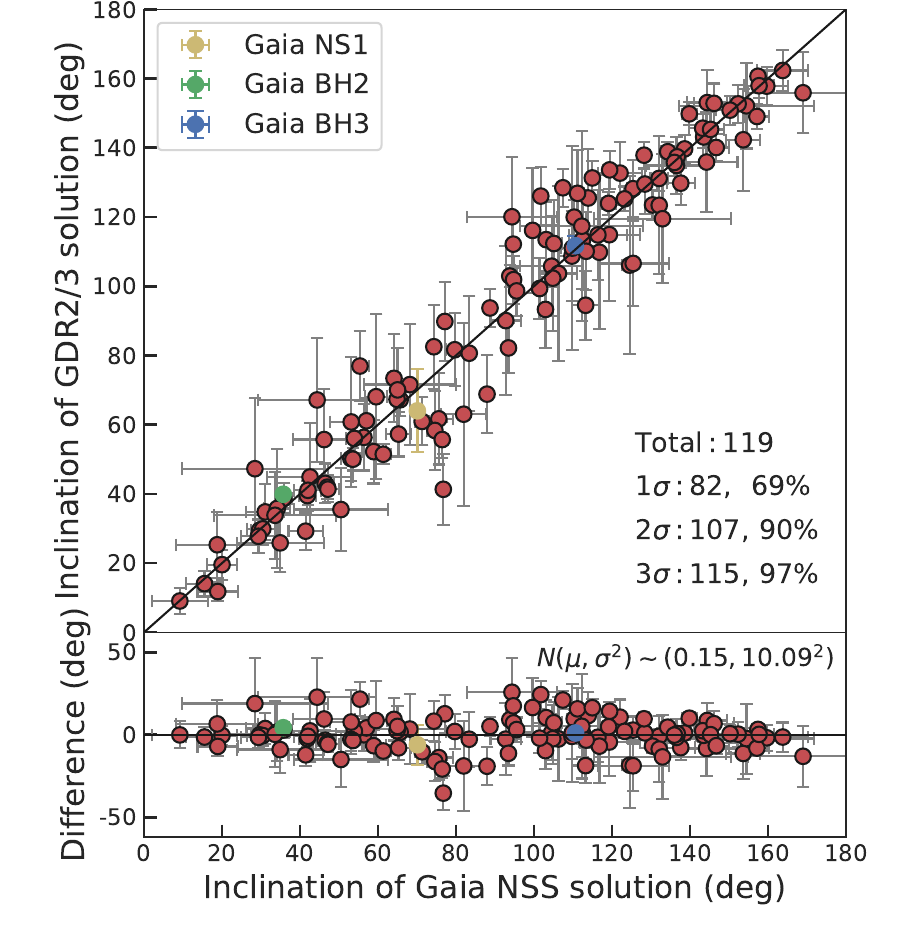}
    \hspace{0.00\textwidth}  
    \includegraphics[width=0.49\textwidth]{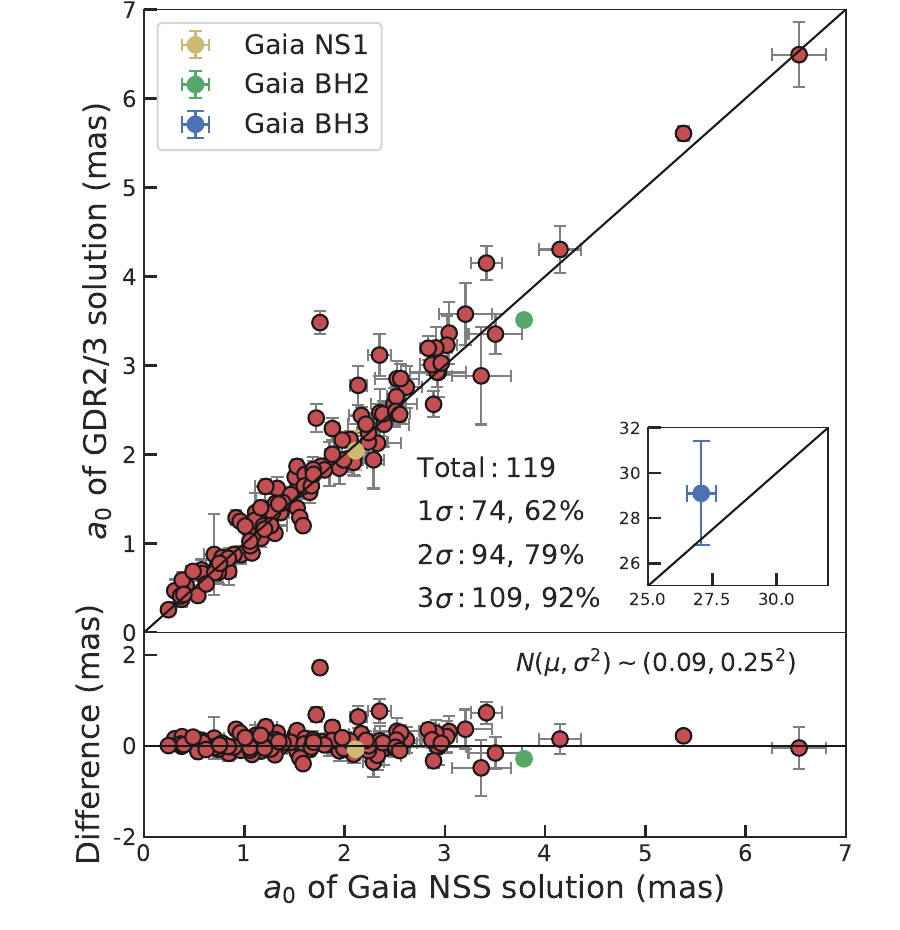}
    \caption{{\bf Comparison of orbital parameters derived from our GDR2/3 method against the {\it Gaia} NSS solutions.} Left panel shows the inclination, while right panel is the photocentric semi-major axis $a_0$, with an inset zooming in on the {\it Gaia} BH3 system. The three benchmark systems were jointly fitted using both RV and GDR2/3 data, whereas the remaining systems are modeled with GDR2/3 astrometry combined with priors from the NSS solution. For {\it Gaia} BH3, we adopted the parameters derived from the {\it Gaia} epoch astrometric data for comparison with our results \citep{GaiaBH3_2024A&A}.
    Statistical metrics are annotated, showing the number and the proportion of sources within $1\sigma$, $2\sigma$, and $3\sigma$ confidence intervals. The Gaussian parameters $N(\mu, \sigma^2)$ for the difference (GDR2/3$-$NSS) distribution are noted in the bottom of each panel. The standard deviation ($10.09^\circ$) of the inclination difference will be quadratically added to our measurement error.}
    \label{compare_NSS_G23.fig}
\end{figure}

Because independent RV measurements are unavailable for the final selected sample, we adopted the NSS-derived values of the orbital parameters that are primarily constrained by RV observations, namely the orbital period $P$, velocity amplitude $K$, eccentricity $e$, argument of periapsis $\omega$, and periapsis epoch $t_{\rm P}$, as external priors. The GDR2/3 absolute astrometry was then used to independently constrain the remaining astrometric orbital parameters, particularly the inclination $i$. In the fitting procedure, we considered nine free parameters: four orbital elements ($P$, $K$, $i$, and $\Omega$) and five astrometric barycentre parameters ($\Delta\alpha_\ast$, $\Delta\delta$, $\Delta\varpi$, $\Delta\mu_{\alpha\ast}$, and $\Delta\mu_\delta$). Gaussian priors were adopted for $P$ and $K$, while $e$, $\omega$, and $t_{\rm P}$ were fixed to the NTBO values to avoid overfitting given the limited astrometric data. All other settings were identical to those adopted for J0008+0128. After the MCMC sampling, systems with poorly constrained inclination measurements ($\sigma_i > 30^\circ$) were excluded from the final sample. 
Figure~\ref{compare_NSS_G23.fig} compares the inclination ($i$) and photocentre semi-major axis ($a_0$) derived from our GDR2/3 method with the values reported in the \textit{Gaia} NSS solutions. We also included three well-constrained systems ({\it Gaia} NS1\citep{2024OJAp....7E..27E}, {\it Gaia} BH2\citep{2023MNRAS.521.4323E} and {\it Gaia} BH3\citep{GaiaBH3_2024A&A}) that host compact stars as benchmarks. Overall, the two methods show quite good agreement, with only a few outliers for $a_0$. The residual distributions for both inclination and $a_0$ are centered near zero, with mean offsets of only $\sim0.15^\circ$ (standard deviation $\sigma=10.09^\circ$) and $\sim0.09$ mas ($\sigma=0.25$\,mas), respectively. This suggests that there is no significant systematic bias between the GDR2/3 method and the \textit{Gaia} NSS solutions.

\subsubsection*{S6. Optical light curves}
\label{sec:lc}

\begin{figure}
    \center
    \includegraphics[width=0.7\textwidth]{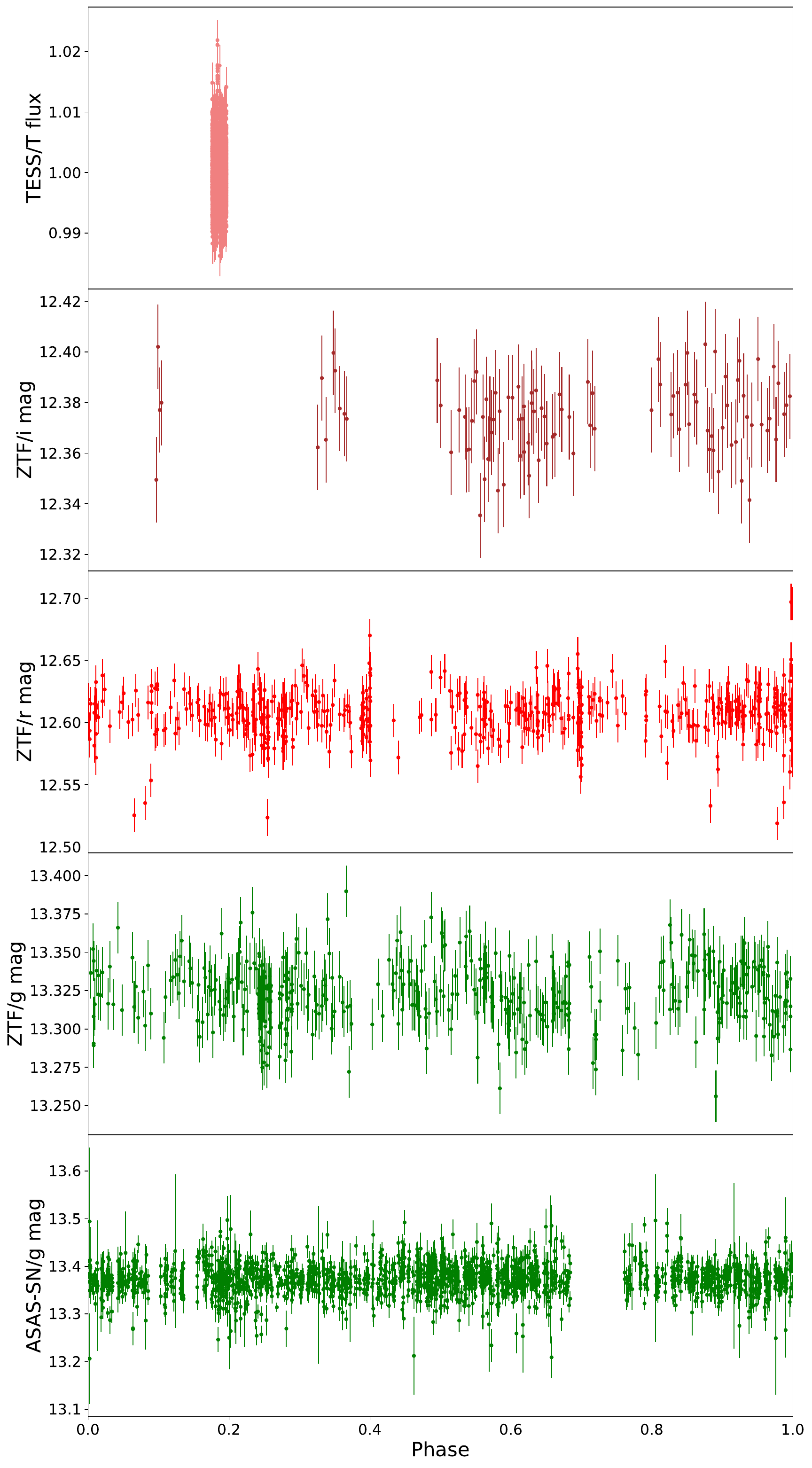}
    \caption{{\bf Observed light curves of J0008+0128 in different bands.} These curves are phased to an orbital period of $P=1217.19$ days.}
    \label{observed_lcs.fig}
\end{figure}

J0008+0128 has been observed by ASAS-SN, ZTF, and TESS.
Therefore, we attempted to determine the orbital period of J0008+0128 using those light curves.
Applying various frequency ranges in the Lomb-Scargle (LS) method \citep{1989ApJ...338..277P}, no clear periodic signal was detected.
Figure \ref{observed_lcs.fig} plots the folded light curves of J0008+0128 using the period of 1217.19 day.
No clear ellipsoidal modulation, which is generally seen in close binaries, is observed in these light curves.

\begin{figure*}
    \center
    \includegraphics[width=0.95\textwidth]{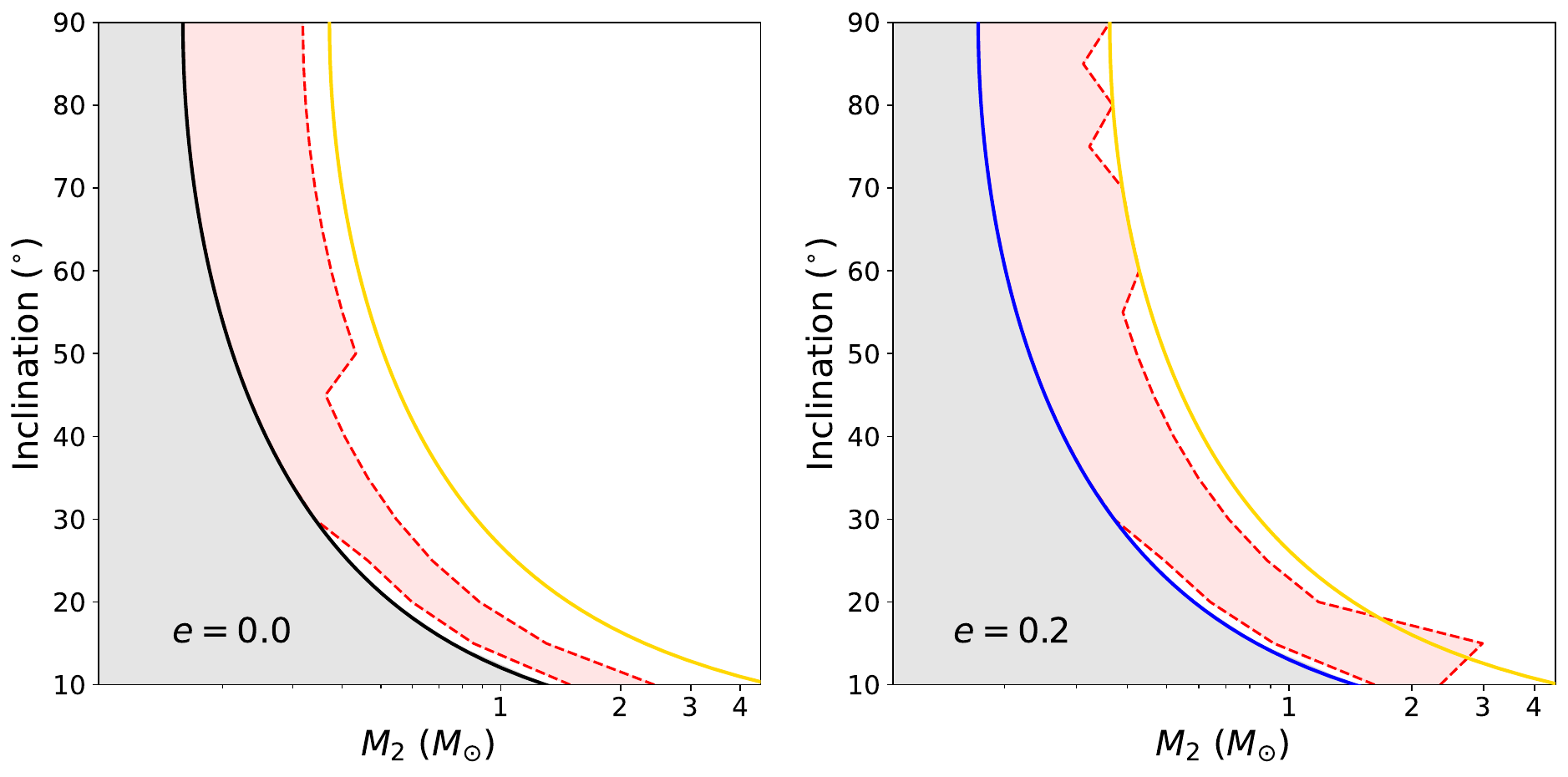}
    \caption{{\bf Mass of the companion star ($M_{\rm 2}$) versus inclination angle.} Black, blue and yellow lines represent mass functions derived for periods of 3, 4, 30 days, respectively. The gray-shaded region corresponds to the parameter range where the visible star fills its Roche lobe, while the red-shaded region highlights the locations where the theoretical amplitude of the TESS light curves exceeds the observed amplitude.
    }
    \label{m2_inclination.fig}
\end{figure*}

However, considering the high precision of TESS in photometric measurements, we can use TESS light curves to exclude the orbital periods that were unsuitable for J0008+0128.
We divided the light curve into 200 bins and then estimated the variation amplitude of the re-binned light curve to be $\Delta f_{\rm obs} \approx 0.0012$.
Assuming the companion is a main-sequence star, we estimated the amplitude of theoretical TESS light curve ($\Delta f_{\rm th}$) using the software Physics of eclipsing binaries \citep{2016ApJS..227...29P, 2018ApJS..237...26H, 2020ApJS..250...34C} (PHOEBE).
For the theoretical light curves, we fixed the mass, radius, and temperature of the visible star at 1.5 $M_{\odot}$, 5.4 $R_{\odot}$, and 5000 K, respectively. 
Additionally, we adopted a range of orbital parameters as below:
\begin{itemize}
    \item Orbital period ($P$): [1, 2, 3, 4, 5, 6, 7, 8, 9, 10, 20, 30, 40, 50, 60, 70, 80, 90, 100, 200, 300, 400, 500, 600, 700, 800, 900] days;
    
    \item Eccentricity ($e$): [0, 0.1, 0.2, 0.3];
    
    \item Inclination angle ($i$): ranging from 10$^{\circ}$ to 90$^{\circ}$ with step of 5$^{\circ}$.
\end{itemize}
Comparing with the observed amplitude $\Delta f_{\rm obs}$, we can exclude orbital parameters that yield a theoretical amplitude $\Delta f_{\rm th}$ larger than $\Delta f_{\rm obs}$.
Figure \ref{m2_inclination.fig} shows the relationship between the mass of companion and inclination for different orbital parameters.
The red-shaded areas indicate the binaries where the theoretical amplitude exceeds the observed amplitude, whereas the gray-shaded regions implies scenarios where the visible star fills its Roche lobe.
It can be seen that the theoretical amplitude of systems with periods shorter than 30 days exceeds the observed amplitude, indicating the period of J0008+0128 is longer than $\approx$30 days.

\subsubsection*{S7. X-ray observation and the upper limit}  
\label{sec:Xray}

The source was not observed by {\it Chandra} or {\it XMM-Newton}, so we performed a Target-of-Opportunity observation at 2025-06-30T01:48:54 (UTC) using the Follow-up X-ray Telescope (FXT) onboard the Einstein Probe (EP). 
Both FXT-A and FXT-B modules were operated in full-frame mode. Over a total exposure time of 9421 s, no X-ray source was detected within 2$^{\prime}$ of J0008+0128 with a significance exceeding 3$\sigma$. 
We used a photometric aperture with a radius of 30$^{\prime\prime}$ to calculated the counts.
An annulus with inner and outer radii of 30$^{\prime\prime}$ and 100$^{\prime\prime}$, respectively, was used to estimate the background emission.
Based on the counts within the photometric aperture and the estimated background, we determined an upper limit on the source counts using a 99\% Bayesian confidence interval \citep{1991ApJ...374..344K}.
This yields upper limits on the count rates of $\approx$ 0.00088 counts s$^{-1}$ for FXT-A and 0.00137 counts s$^{-1}$ for FXT-B, respectively.

To estimate the flux upper limit, we used the {\tt\string fakeit} tool in the Xspec package to simulate absorbed power-law spectra, incorporating the instrumental response. 
The hydrogen column density $N_{\rm H}$ was estimated using the linear relation between $N_{\rm H}$ and the reddening: $N_{\rm H}=f\times10^{21}$ cm$^{-2}$ $E{\rm (B-V)}$.
Previous studies \citep{1995A&A...293..889P,2009MNRAS.400.2050G,2011A&A...533A..16W,2015ApJ...809...66V,2016ApJ...826...66F,2017MNRAS.471.3494Z} suggested that the value of the factor $f$ ranges from 5.55 to 8.90 when adopting an extinction coefficient of $R_{V}=3.1$.
In this work, we utilized results from Predehl and Schmitt \citep{1995A&A...293..889P}, which provided a value of $f=5.55$ based on ROSAT observations of interstellar medium.
The simulations assumed a hydrogen column density of $N_\mathrm{H}=5.55\times10^{19} \mathrm{cm}^{-2}$ and photon index of $\tau=1.5$ and $\tau=2$. 
The unabsorbed flux upper limits in 0.5-10 keV for FXT-A are $2.88\times10^{-14} \rm erg\, s^{-1}\, cm^{-2}$ and $1.98\times10^{-14} \rm erg\, s^{-1}\, cm^{-2}$ for $\tau=1.5$ and $\tau=2$ repectively. 
And the unabsorbed flux upper limits for FXT-B are $3.97\times10^{-14} \rm erg\, s^{-1}\, cm^{-2}$ and $2.71\times10^{-14} \rm erg\, s^{-1}\, cm^{-2}$ repectively.
Taking the distance of 1464.56 pc, we further calculated the upper limit X-ray luminosities of 8.8$\times$10$^{30}$ $\rm erg\, s^{-1}$ ($\tau=1.5$) and 6.0$\times$10$^{30}$ $\rm erg\, s^{-1}$ ($\tau=2$).

To check whether the binary nature of J0008+0128 (a giant star plus an NS) is consistent with the observed X-ray upper limit, we examined three possible sources of X-ray emission:

(1) Wind accretion by the NS. We used the Bondi–Hoyle–Lyttleton (BHL) accretion to estimate the expected X-ray flux assuming a spherically symmetric wind from the visible star,
\begin{equation}
\label{BHL}
    \dot{M}_{{\rm BHL}}	=\frac{G^{2}M_{{\rm 2}}^{2}\dot{M}_{{\rm wind}}}{v_{{\rm wind}}^{4}a^{2}}.
\end{equation}
The wind mass-loss rate ($\dot{M}_{{\rm wind}}$) can be calculated following Reimers \citep{1975MSRSL...8..369R},
\begin{equation}
    \label{wind_mdot}
    \dot{M}_{{\rm wind}}=4\times 10^{-13}\eta_{R}M_{\odot}\,{\rm yr}^{-1}\left(\frac{L_{1}}{L_{\odot}}\right)\left(\frac{R_{1}}{R_{\odot}}\right)\left(\frac{M_{1}}{M_{\odot}}\right)^{-1}.
\end{equation}
We applied $\eta_R=0.1$ following Choi et al. \citep{2016ApJ...823..102C}. 
The velocity of wind ($v_{{\rm wind}}$) can be expressed by the escape velocity at the stellar surface:
\begin{equation}
    \label{vwind}
    v_{{\rm wind}}=\sqrt{\frac{2GM_{1}}{R_{1}}}.
\end{equation}

Assuming isotropic emission, we estimated the expected X-ray flux using equations \ref{BHL}, \ref{wind_mdot} and \ref{vwind} following
\begin{equation}
\label{fx}
F_{\rm BHL}=\frac{\eta_{X}\dot{M}_{{\rm BHL}}c^{2}}{4\pi d^{2}},
\end{equation}
where $\eta_{X}$ is radiative efficiency, and $d$ is the distance of J0008+0128.
Using the fitting function obtained from Xie and Yuan \citep{2012MNRAS.427.1580X}, we estimated the value of $\eta_{X}$.
Consequently, the expected X-ray flux of J0008+0128 is about 4.2$\times 10^{-19}\ \rm erg\, s^{-1}\, cm^{-2}$, resulting in an X-ray luminosity of 1.1$\times$10$^{26}$ $\rm erg\, s^{-1}$, far below the detection limit of current X-ray telescopes.

In addition, we also adopted a modified scaling factor of $\eta=0.172$ (replacing $\eta_R=0.1$) \citep{2005ApJ...630L..73S,2015MNRAS.448..502M} for red giant branch stars to estimate the wind mass-loss rate. This results in an X-ray luminosity of 2.6$\times$10$^{26}$ $\rm erg\, s^{-1}$, further indicating that the wind accretion luminosity is very low.

(2) Thermal emission from the NS.
Thermal blackbody emission from the surface of an isolated NS can also produce detectable X-ray luminosity \citep{2022MNRAS.509.1217R,2025arXiv250112143R}. 
The well-known ``Magnificent Seven'' isolated NS have X-ray luminosities of $L_X \sim 10^{31}$ erg s$^{-1}$.
As an NS cools, its surface temperature and X-ray luminosity gradually decline, with the cooling proceeding more rapidly for higher-mass NSs \citep{2004ARA&A..42..169Y,2025AN....34640110H}.
For example, a 2.2 $M_{\odot}$ NS can cool to an X-ray luminosity of $\sim10^{31}$ $\rm erg\, s^{-1}$, comparable to the upper limit of J0008+0128, within only $\sim$ 10$^3$ to 10$^5$ years, depending on the assumed equation of state  \citep{2020MNRAS.496.5052P,2024PhRvD.109l3018D}. 
Although the age of the NS in J0008+0128 cannot be determined, the short cooling timescale makes the presence of a young, supermassive NS unlikely. Consequently, its thermal emission should lie well below the observed X-ray upper limit.

\begin{figure}
    \center
    \includegraphics[width=0.7\textwidth]{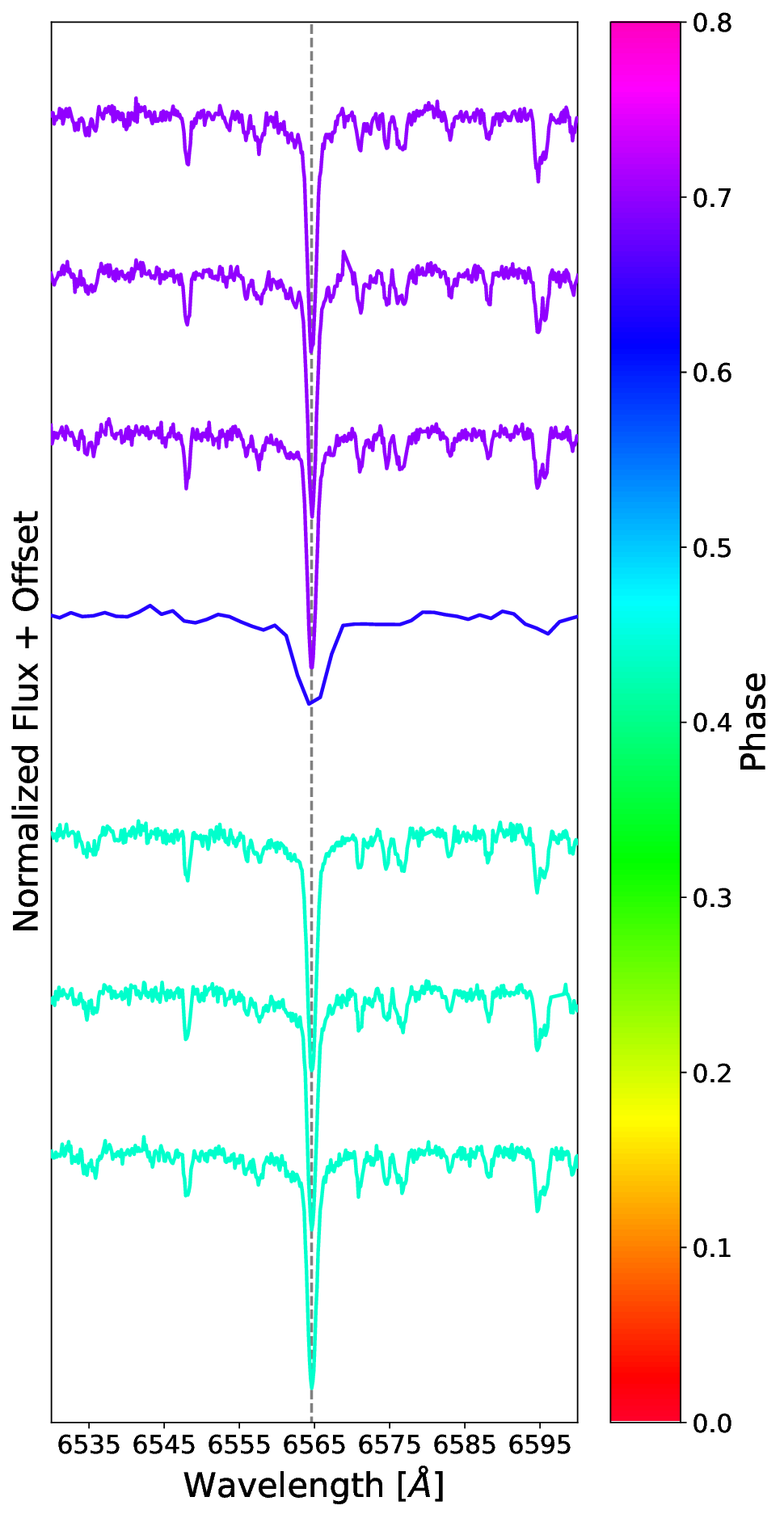}
    \caption{{\bf Observed $H_{\alpha}$ line profiles of J0008+0128.} The spectra from LAMOST LRS and MRS observations are plotted at different phases.}
    \label{Ha.fig}
\end{figure}

\begin{figure}
    \center
    \includegraphics[width=0.99\textwidth]{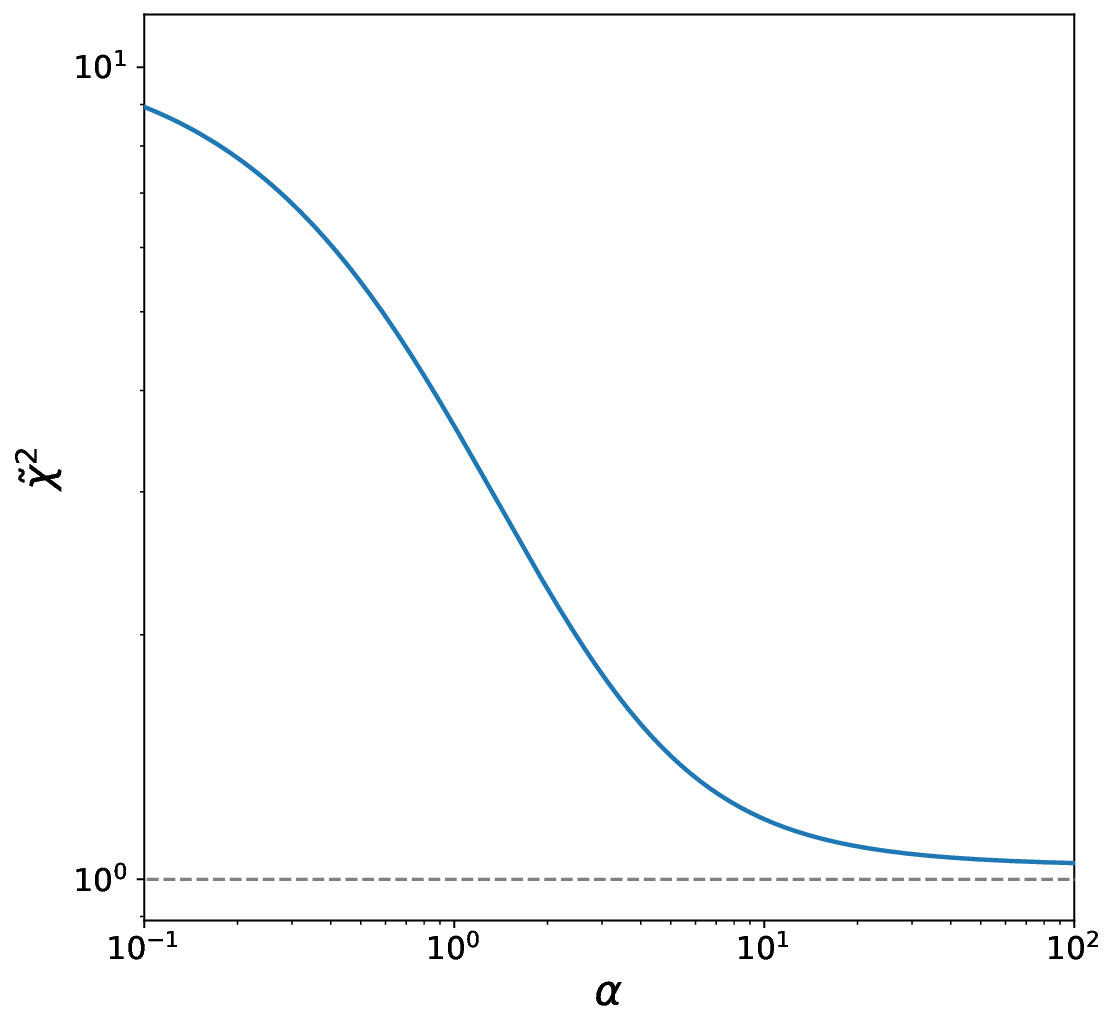}
    \caption{{\bf Light ratio $\alpha$ between the two components derived from spectral disentangling, for $q = 0.79$}.
    The dash line marks the $\tilde{\chi}^2=1$.}
    \label{luminosity_ratio.fig}
\end{figure}

\begin{figure*}
    \center
    \includegraphics[width=0.9\textwidth]{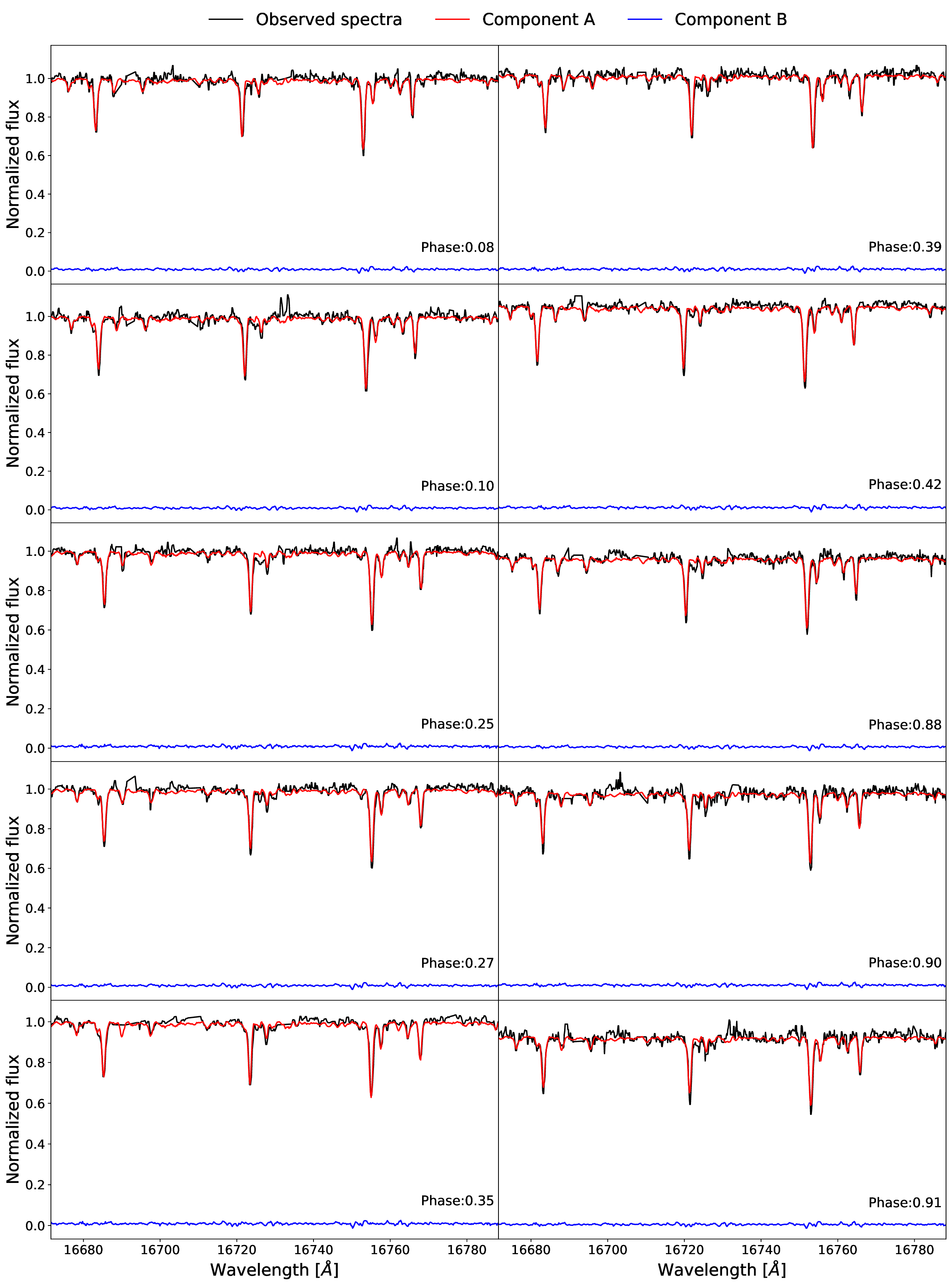}
    \caption{{\bf 
    Phase-resolved spectral disentangling for $q = 0.79$.} Red lines mark the reconstructed spectra of the visible star, while blue lines represent the other component in each spectra, plotted with a light ratio of $\alpha=1/100$ (see Figure \ref{luminosity_ratio.fig}). Black lines are the observed spectra.}
    \label{disentangling_results.fig}
\end{figure*}

(3) Coronal X-ray emission from the visible giant star. 
Late-type giants with high magnetic activity can produce substantial X-ray emission, with luminosities reaching $\sim10^{30}$--$10^{31}$ $\rm erg\, s^{-1}$ \citep{1996A&A...309L..51H,2020ApJ...902..114W,2024A&A...688A...9S}. 
Such giants typically also exhibit strong ultraviolet emission, as well as prominent
Ca H\&K and $H_{\alpha}$ emission lines \citep{2008ApJ...687.1264M,2013MNRAS.431.2063S}. 
However, in the case of J0008+0128, the observed NUV emission is nearly consistent with the photospheric level predicted by templates (Figure \ref{fig1.fig}); the CaII H \& K lines show quite weak emission; the $H_{\alpha}$ line exhibits no emission feature (Figure \ref{Ha.fig}). All these suggest the giant in J0008+0128 has very low magnetic activity, and consequently, low X-ray emission.

Consequently, all scenarios discussed here are consistent with the observed X-ray upper limit.

\subsubsection*{S8. FAST observation}
\label{sec:fast}

The Five-hundred-meter Aperture Spherical Telescope (FAST) \citep{2006ScChG..49..129N} was employed to search for radio pulses from the invisible companion. 
Initially, two tracking observations, each lasting 1800 seconds, were conducted on February 6th and 8th, 2025. 
The data were sampled at a time resolution of 49.152 microseconds, with 2048 frequency channels covering a 500 MHz bandwidth centered at 1.25 GHz. Assuming a pulse duty cycle of 10\%, these integrations yielded a sensitivity limit of approximately 2 $\mu$Jy. We employed the Galactic Plane Pulsar Snapshot Survey (GPPS) pulsar searching pipeline \citep{2021RAA....21..107H} for the periodic search. This data processing pipeline is based on PRESTO (Ransom, S. 2011, PRESTO: PulsaR Exploration and Search TOolkit), and during the search, a DM range of 0.5--3057 pc cm$^{-3}$ was covered.
However, we failed to detect any pulsed signals. 
Subsequently, two additional observations with longer integration times of 3600 seconds were carried out on March 9th and April 14th, 2025. 
Two weak pulsar candidates were identified from each observation: one with a spin period of 1.88 ms and a dispersion measure (DM) of 155.5 ${\rm pc} \ {\rm cm}^{-3}$, and the other with a spin period of 3.27 ms and a DM of 470.6 ${\rm pc} \ {\rm cm}^{-3}$ as from the GPPS pulsar searching pipeline.
%
Unfortunately, neither candidate was re-detected in the other observation. The visible star's distance of 1464.56 pc implies a DM of 15.23 ${\rm pc} \ {\rm cm}^{-3}$ according to the YWM16 electron density model \citep{2017ApJ...835...29Y}, which deviates significantly from the DMs of the two candidates. 
Consequently, no radio pulsar counterpart was found for the system.

If the invisible star in the system is indeed a massive radio pulsar, the non-detection could be attributed to either a narrow radio emission beam that does not sweep across Earth or very weak pulse emission that remains undetectable even with the most sensitive radio telescope, FAST.
As established earlier, the visible star is non-degenerate. To our knowledge, there are two types of systems where a pulsar’s companion can be a non-degenerate star \citep[e.g.][]{2023pbse.book.....T}. 
For the first type, the young pulsars have not accreted matter from their main-sequence companions. 
These systems generally have wide and non-circular orbits (eccentricity $e>0.05$, orbital period P$_{\rm orb}>80$ days), with main-sequence companions having masses larger than 1.0 $M_{\odot}$, and the pulsars are non-recycled \citep{2025RAA....25a4003W}. 
For the second type, millisecond pulsars have accreted matter from companions via Roche lobe overflow (RLOF), where the companion star evaporation leads to orbital detachment. 
These systems typically exhibit close and circular orbits ($e<0.001$, P$_{\rm orb}<1.1$ day) and have companions with masses of 0.1 $M_{\odot}<m<0.52 M_{\odot}$. 
If the invisible star in J0008+0128 is a pulsar, the system matches the first type in those key aspects, i.e., mass, eccentricity, and orbital period.

\subsubsection*{S9. Searching for similar systems in POSYDON population}
\label{sec:formation_history}

To assess the likelihood of forming such a system from an isolated binary, we examined the population of solar-metallicity binaries simulated with the \texttt{POSYDON} (version 1) binary population synthesis framework \citep{Fragos2023,Andrews2024}. In this population, two million binaries were evolved with initial conditions motivated by observations: primary masses drawn from a Kroupa IMF ($M_1 = 7$–120 $M_\odot$), secondary masses sampled uniformly in $q = M_2/M_1 = [0,1]$ down to $M_2 = 1\,M_\odot$, and initial orbital periods from 0.35 to 3162 days following Sana et al. \citep{Sana2013}. Binary evolution was modeled self-consistently using detailed \texttt{MESA} grids \citep{Paxton2011,Paxton2013,Paxton2015,Paxton2018,Paxton2019,Jermyn2023} to evolve both stars (unless one is treated as a point mass), including phases of stable mass transfer. Donor and accretor are evolved simultaneously during Roche lobe overflow, enabling consistent treatment of mass transfer, stellar response, and orbital evolution. Common envelope (CE) evolution and supernova explosion were treated within the \texttt{POSYDON} framework using analytic prescriptions adopted for this study: the CE phase is modeled with the standard $\alpha_{\rm CE}$–$\lambda_{\rm CE}$ formalism (assuming $\alpha_{\rm CE} = 1$ and computing $\lambda_{\rm CE}$ from internal structure), and compact object formation follows the core-collapse SN prescription proposed by Patton et al. \citep{Patton2020} and the electron-capture SN prescription presented by Podsiadlowski et al. \citep{Podsiadlowski2004}. NS kicks were drawn from Maxwellian distributions with $\sigma_{\rm CCSN}=265\,\mathrm{km\,s^{-1}}$ and $\sigma_{\rm ECSN}=20\,\mathrm{km\,s^{-1}}$, and the NS mass was capped at $2.5\,M_\odot$. Other modeling details can be found in Rocha et al. \citep{Rocha2024}. Although this population was originally constructed to study Be X-ray binaries, its initial conditions are broadly representative of binaries with an intermediate- to high-mass primary (i.e., NS or BH progenitors) and a lower-mass secondary. The parameter space fully encompasses systems like J0008+0128 and is thus suitable for evaluating the expected formation rate of wide NS binaries.

\begin{figure*}
    \center
    \includegraphics[width=0.95\textwidth]{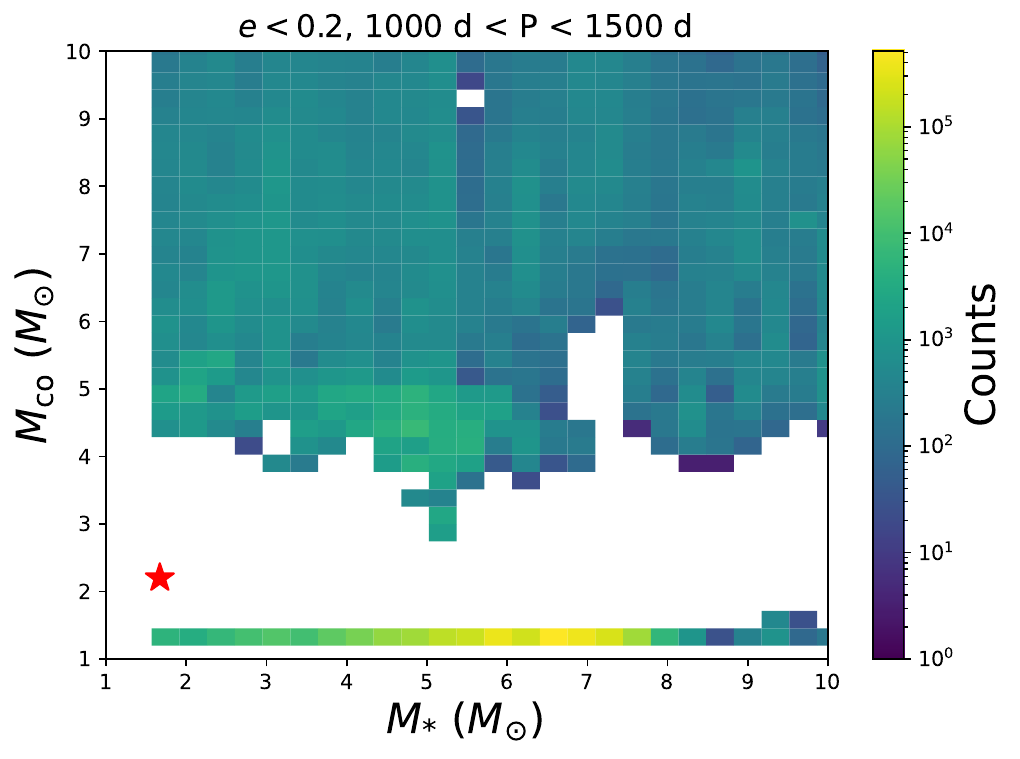}
    \caption{{\bf Number distribution of compact binaries from POSYDON.} The sample is selected based on an eccentricity of $e < 0.2$ and orbital periods between 1000 and 1500 days. $M_{*}$ denotes the mass of the visible star, and $M_{\rm co}$ represents the mass of the compact object. A red star marks the best-fitting location of our system.}
    \label{posydon_M1_M2.fig}
\end{figure*}

\begin{figure*}
    \center
    \includegraphics[width=1\textwidth]{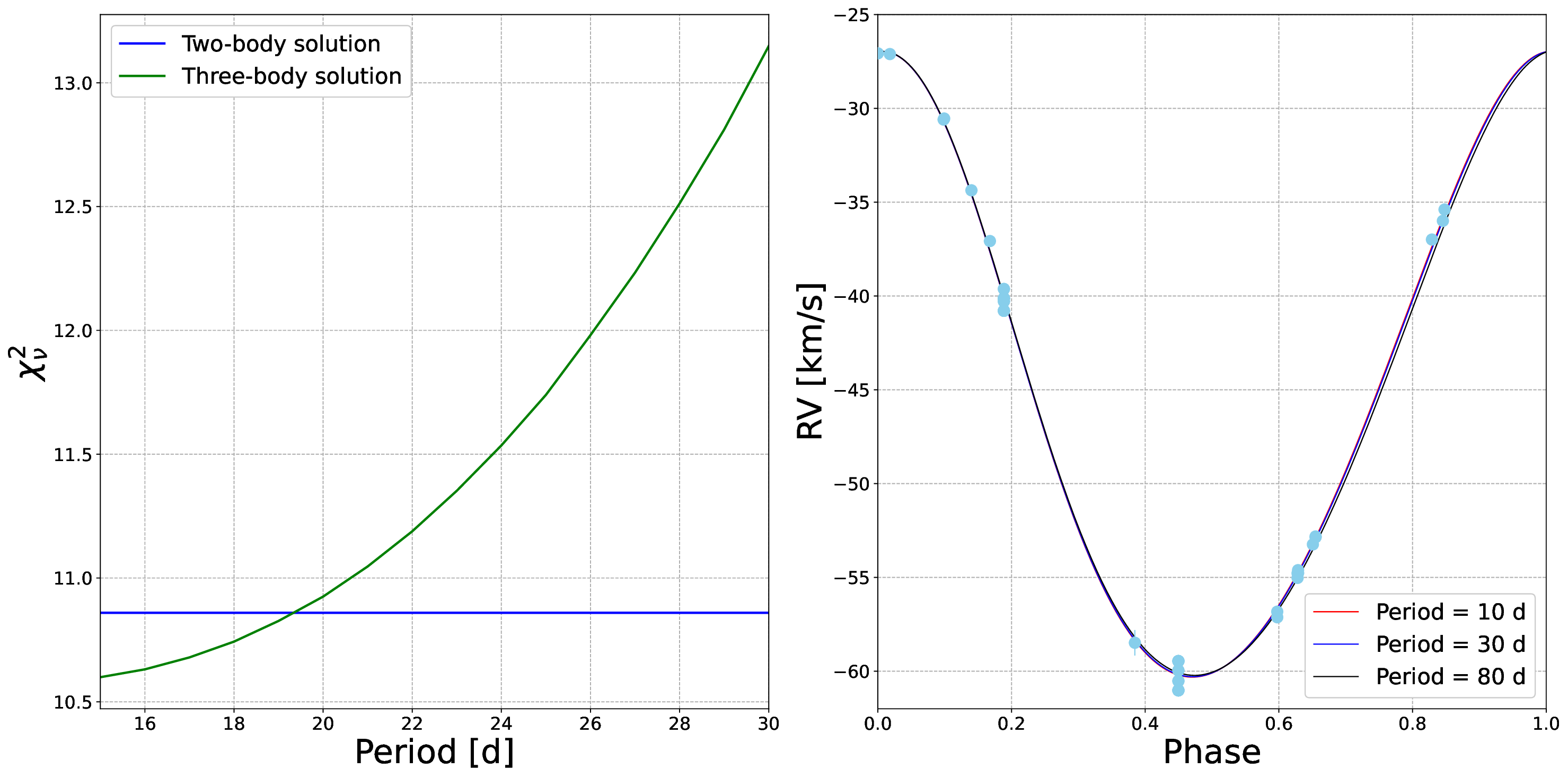}
    \caption{{\bf Comparison between the two-body solution and the hierarchical triple model.} Left panel: Reduced--$\chi^2$ ($\chi^2_{\nu}$) values for different inner orbital periods. The blue line shows the two-body solution obtained with {\it The Joker}. The green line represents the results from a hierarchical triple model, assuming an inner binary composed of two compact objects, each with a mass of $1 M_{\odot}$. Right panel: Simulated RV curves for inner orbital periods of 10, 30, and 80 days obtained with the {\it rebound} integrator.}
    \label{2n_3n.fig}
\end{figure*}

We performed a systematic search over the full binary population, scanning the entire evolutionary history of all simulated systems for epochs that satisfy a deliberately broad parameter window motivated by the observed properties of J0008+0128. 
We identified the binary systems with properties similar to J0008+0128 with following conditions: 
orbital periods in the range $1000$--$1500\,\mathrm{d}$, eccentricities $e < 0.2$, primary masses between 1.8 and $3.0\,M_\odot$, and secondary masses between 1.0 and $2.9\,M_\odot$ (corresponding to the visible star). 
These criteria are wider than the observational constraints, in order to avoid excluding potential progenitors a priori.
However, we did not identify any analogs to J0008+0128 in this fiducial population (Figure \ref{posydon_M1_M2.fig}).
This result indicates that binaries resembling J0008+0128 cannot be formed under standard assumptions within the POSYDON population synthesis framework.

The observed configuration of J0008+0128, which consists of a red giant of 1.2--2.7 $M_{\odot}$ orbiting a compact object of 1.8--2.8 $M_{\odot}$ with an orbital period near 1200 days and eccentricity around 0.1, may require a combination of extreme but physically plausible parameters, such as a high NS birth mass, minimal supernova kick, and unusually efficient CE ejection.

The progenitor system likely began with a massive primary star ($\gtrsim 8\,M_\odot$) and a lower-mass secondary ($\sim$2--3 $M_\odot$), with an initial orbital separation of several hundreds to a thousand solar radii. As the primary evolved off the main sequence and expanded, it would have filled its Roche lobe and initiated mass transfer. Depending on the mass ratio and orbital separation, this phase may have been stable or may have led to a CE episode. In either case, the envelope of the primary would have been stripped, leaving behind a short-lived helium star. 

This helium star subsequently underwent core collapse to form an NS with a mass of 1.8--2.8 $M_{\odot}$. The current low orbital eccentricity implies that the natal supernova kick was either intrinsically small or oriented such that the binary remained bound with only modest orbital perturbation. This aligns with growing evidence for low-kick NS formation channels \citep{Valli2025}.

Today, the visible star, the original secondary, has evolved off the main sequence into a red giant with a mass of $\sim$ 1.2--2.7 $M_{\odot}$, while the compact companion is most likely a massive NS. Given the current separation, the red giant may eventually fill its Roche lobe and initiate a phase of late Case C mass transfer, potentially evolving into a wide NS–WD binary in the future.

Although it is in principle possible that the NS formed without ever interacting with its companion (during the whole evolution), such a scenario would require a combination of a low natal kick, unusually high birth mass, and fine-tuned initial orbital parameters. It also remains unclear whether incorporating a low-kick prescription into binary population synthesis can simultaneously reproduce the full diversity of observed systems. We therefore consider this to be less likely than an evolutionary channel involving at least one phase of mass transfer or CE evolution. Like {\it Gaia} NS1 ($P=730.8\pm1.2$ day, $e=0.122\pm0.002$, $M_{\rm NS} =1.90\pm0.04\ M_{\odot}$) \citep{2024OJAp....7E..27E} and G3425 ($P=879.11^{+3.22}_{-2.57}$ day, $e=0.05^{+0.01}_{-0.01}$, $M_{\rm BH} =3.58^{+0.80}_{-0.47}\ M_{\odot}$) \citep{2024NatAs...8.1583W}, such wide and low-eccentricity orbits are thought to be more likely formed through a CE phase. However, this scenario requires a significantly higher CE ejection efficiency than typically assumed.

\clearpage

\begin{table}
\caption{\bf Barycentric-corrected RV values of J0008+0128. \label{rvdata.tab}}
\centering
\setlength{\tabcolsep}{1.2mm}
\renewcommand\arraystretch{0.8}
\begin{center}
\begin{tabular}{cccc}
\hline\noalign{\smallskip}
BJD & RV & S/N & Res. \\
(day) & (km/s) &  &  \\
\hline\noalign{\smallskip}
2456644.62918 & $-27.07\pm0.42$ & 100.45 & APOGEE/DR17\\
2456666.58257 & $-27.11\pm0.39$ & 100.61 & APOGEE/DR17\\
2456848.94608 & $-37.07\pm0.42$ & 107.50 & APOGEE/DR17\\ 
2457654.83847 & $-36.99\pm0.34$ & 79.34 & APOGEE/DR17\\ 
2457674.74723 & $-36.00\pm0.45$ & 7.69 & APOGEE/DR17\\   
2457677.76332 & $-35.39\pm0.34$ & 92.64 & APOGEE/DR17\\ 
2458033.75983 & $-34.37\pm0.34$ & 72.26 & APOGEE/DR17\\ 
2458092.94713 & $-40.76\pm0.34$ & 78.82 & LAMOST/MRS $r$-band \\  
2458092.94713 & $-39.59\pm0.33$ & 39.60 & LAMOST/MRS $b$-band \\   
2458092.96379 & $-40.76\pm0.34$ & 78.29 & LAMOST/MRS $r$-band \\  
2458092.96379 & $-40.09\pm0.34$ & 39.66 & LAMOST/MRS $b$-band \\  
2458092.97976 & $-40.09\pm0.34$ & 32.88 & LAMOST/MRS $b$-band \\  
2458092.97976 & $-40.26\pm0.34$ & 69.62 & LAMOST/MRS $r$-band \\  
2458411.07275 & $-61.02\pm0.30$ & 63.99 & LAMOST/MRS $r$-band \\   
2458411.07275 & $-59.96\pm0.31$ & 33.76 & LAMOST/MRS $b$-band \\  
2458411.08941 & $-61.02\pm0.29$ & 65.27 & LAMOST/MRS $r$-band \\  
2458411.08941 & $-59.46\pm0.31$ & 35.78 & LAMOST/MRS $b$-band \\  
2458411.10538 & $-60.52\pm0.29$ & 61.97 & LAMOST/MRS $r$-band \\  
2458411.10538 & $-59.46\pm0.31$ & 31.40 & LAMOST/MRS $b$-band \\   
2459201.61242 & $-30.60\pm0.35$ & 130.32 & APOGEE/SDSS-V \\  
2459202.58266 & $-30.54\pm0.32$ & 105.70 & APOGEE/SDSS-V \\  
2459550.28464 & $-58.18\pm0.91$ & 51.23 & LAMOST/LRS \\    
2459846.84877 & $-54.78\pm0.34$ & 86.12 & APOGEE/SDSS-V \\  
2459846.88674 & $-55.02\pm0.33$ & 102.83 & APOGEE/SDSS-V \\ 
2459847.78396 & $-54.61\pm0.30$ & 62.55 & APOGEE/SDSS-V \\   
2459847.81121 & $-54.77\pm0.31$ & 78.47 & APOGEE/SDSS-V \\  
2459874.80739 & $-53.24\pm0.37$ & 53.76 & APOGEE/SDSS-V \\  
2459879.71614 & $-52.82\pm0.35$ & 59.46 & APOGEE/SDSS-V \\  
2459879.73612 & $-52.86\pm0.39$ & 55.90 & APOGEE/SDSS-V \\ 
2461028.34698 & $-56.60\pm0.05$ & 10.55 & NOT \\ 
2461028.36546 & $-56.90\pm0.06$ & 8.20 & NOT \\ 
\noalign{\smallskip}\hline
\end{tabular}
\end{center}
\end{table}

\begin{table}
\caption{{\bf Estimation of stellar parameters with different methods for J0008+0128.} The LASP results are from the LAMOST DR11 parameter catalog. For each method, the first row lists the atmospheric parameters adopted from the catalogs, while the second row gives the corrected values.\label{atmo_pars.tab}}
\centering
\renewcommand{\arraystretch}{1.2}
 \begin{tabular}{llllll}
\hline\noalign{\smallskip}
Methods & $T_{\rm eff}$ (K) & log$g$ & [Fe/H] & Catalog$^a$\\
\hline\noalign{\smallskip}
\multicolumn{5}{c}{APOGEE/DR17}
\\\hline
\multirow{2}{*}{ASPCAP} & $4955{\pm 10}$ & $3.15{\pm 0.03}$ & $-0.25{\pm 0.01}$ & (1) \\
& $4964{\pm 40}$ & $3.14{\pm 0.07}$ & $-0.24{\pm 0.05}$ & \\
\hline
\multicolumn{5}{c}{APOGEE/SDSS-V}
\\\hline
\multirow{2}{*}{ASPCAP} & $4913{\pm 13}$ & $3.20{\pm 0.03}$ & $-0.23{\pm 0.01}$ & (2) \\
& $4984{\pm 40}$ & $3.19{\pm 0.06}$ & $-0.22{\pm 0.06}$ &  \\
\hline
\multicolumn{5}{c}{LAMOST/LRS}
\\\hline
\multirow{2}{*}{LASP} & $5009{\pm 26}$ & $3.35{\pm 0.04}$ & $-0.23{\pm 0.02}$ & (3) \\
& $4993{\pm 41}$ & $3.27{\pm 0.08}$ & $-0.22{\pm 0.06}$ & \\
\multirow{2}{*}{DD-Payne} & $4986{\pm 30}$ & $3.29{\pm 0.07}$ & $-0.29{\pm 0.03}$ & (4) \\
& $4986{\pm 30}$ & $3.30{\pm 0.10}$ & $-0.22{\pm 0.07}$ & \\
\hline
\multicolumn{5}{c}{LAMOST/MRS}
\\\hline
\multirow{2}{*}{LASP} & $4992{\pm 22}$ & $3.26{\pm 0.05}$ & $-0.30{\pm 0.03}$ & (5) \\
& $4998{\pm 63}$ & $3.20{\pm 0.10}$ & $-0.28{\pm 0.06}$ & \\
\noalign{\smallskip}\hline
\end{tabular}\\
$^a$ (1) APOGEE DR17 stellar parameter catalog from ASPCAP \citep{2022ApJS..259...35A}. 
(2) APOGEE/SDSS-V stellar parameter catalog from ASPCAP\footnote{https://data.sdss.org/sas/dr19/env/MWM\_ASTRA/0.6.0/summary/}. 
(3) LAMOST DR11 LRS stellar parameter catalog\footnote{http://lamost.org/dr11/catalogue}. 
(4) LAMOST DR11 LRS stellar parameter catalog from the DD-Payne method \citep{2025ApJS..279....5Z}. 
(5) LAMOST DR11 MRS stellar parameter catalog\footnote{http://lamost.org/dr11/catalogue}.
\end{table}

\begin{table*}
\caption{\bf RV fitting results from the {\it The Joker} and {\it Gaia} DR3 NTBO table. \label{keplerorbit.tab}}
\centering
\renewcommand{\arraystretch}{1.5}
\setlength{\tabcolsep}{0.5mm}
 \begin{tabular}{ccccccccc}
\hline\noalign{\smallskip}
Methods & $P$ & $e$ & $\omega$ & $M{\rm 0}$ & $K_{\rm 1}$ & $\nu{\rm 0}$ & $f$(m) \\
 & (day) & & (rad) & (rad) & (km/s) & (km/s) & ($M_{\odot}$) \\
\hline\noalign{\smallskip}
Our fit & $1217.19^{+0.58}_{-0.51}$ & $0.08^{+0.01}_{-0.01}$ & $0.62^{+0.02}_{-0.03}$ & $0.57^{+0.02}_{-0.02}$ & $16.75^{+0.03}_{-0.03}$ & $-44.70^{+0.02}_{-0.02}$ & $0.587^{+0.004}_{-0.004}$ \\
{\it Gaia} DR3 & $1199.39\pm267.39$ & $0.20\pm0.27$ & -- & -- & $16.62\pm1.56$ & $-45.46\pm3.15$ & $0.537\pm0.387$ \\
\noalign{\smallskip}\hline
\end{tabular}
\end{table*}

 \begin{table*}
 \caption{{\bf Flux ratios of the companion to the visible star for different values of $M_{1}$ and orbital inclination.} The subscripts $G$, $BP$, $RP$, $J$, $H$, and $K_{s}$ denote the photometric bands of {\it Gaia} and {\it 2MASS}, respectively. \label{light_ratios.tab}}
 \begin{center}
\renewcommand{\arraystretch}{0.9}
  \begin{tabular}{cccccccc}
 \hline\noalign{\smallskip}
$M_{1}$ & $M_{2}$  & $f_{G}$ & $f_{BP}$ & $f_{RP}$ & $f_{J}$ & $f_{H}$ & $f_{K_{s}}$ \\
($M_{\odot}$) & ($M_{\odot}$) &  &  &  &  &  &  \\
 \hline\noalign{\smallskip}       
\multicolumn{8}{c}{$i=90^{^\circ}$}
\\\hline
0.60 & 1.27 & 0.22 & 0.26 & 0.17 & 0.13 & 0.09 & 0.09 \\
0.80 & 1.43 & 0.34 & 0.42 & 0.25 & 0.18 & 0.13 & 0.12 \\
1.00 & 1.57 & 0.51 & 0.67 & 0.36 & 0.23 & 0.16 & 0.15 \\
1.20 & 1.70 & 0.67 & 0.92 & 0.45 & 0.28 & 0.19 & 0.18 \\
\hline
\multicolumn{8}{c}{$i=80^{^\circ}$}
\\\hline
0.60 & 1.31 & 0.25 & 0.30 & 0.19 & 0.14 & 0.10 & 0.10 \\
0.80 & 1.47 & 0.39 & 0.50 & 0.28 & 0.20 & 0.14 & 0.13 \\
1.00 & 1.61 & 0.54 & 0.73 & 0.38 & 0.25 & 0.17 & 0.16 \\
1.20 & 1.75 & 0.70 & 0.99 & 0.47 & 0.29 & 0.19 & 0.18 \\
\hline
\multicolumn{8}{c}{$i=70^{^\circ}$}
\\\hline
0.60 & 1.43 & 0.34 & 0.42 & 0.25 & 0.18 & 0.13 & 0.12 \\
0.80 & 1.60 & 0.53 & 0.70 & 0.37 & 0.24 & 0.17 & 0.16 \\
1.00 & 1.75 & 0.70 & 0.99 & 0.47 & 0.29 & 0.19 & 0.18 \\
1.20 & 1.89 & 0.89 & 1.29 & 0.56 & 0.32 & 0.21 & 0.20 \\
\noalign{\smallskip}\hline
\end{tabular}
\end{center}
\end{table*}

\begin{table*}
\caption{\bf Information of other neutron stars and candidates. \label{other_ns.tab}}
\centering
\setlength{\tabcolsep}{0.05cm}
\renewcommand\arraystretch{0.9}
\begin{center}
\begin{tabular}{lcccccc}
\hline\noalign{\smallskip}
Name & $P$ & $e$ & $M_{\rm c}$ & $M_{\rm NS}$ & Type & References \\
 & (day) &  & ($M_{\odot}$) & ($M_{\odot}$) &  \\
\hline\noalign{\smallskip}
J1141-6545 & 0.1976509593(1) & 0.171884(2) & 1.02(1) & 1.27(1) & NS+WD & (1) \\   
J1906+0746 & 0.16599304686(11) & 0.0852996(6) & 1.322(11) & 1.291(11) & DNS & (2) \\   
B1534+12 & 0.420737298881(2) & 0.27367740(4) & 1.3455(2) & 1.3330(2) & DNS & (3) \\
J0737-3039A & 0.102251561(8) & 0.087793(8) & 1.2490(72) & 1.3381(7) & DNS & (4) \\
J1756-2251 & 0.319633898(2) & 0.1805694(2) &  1.230(7) & 1.341(7) & DNS & (5) \\
B2127+11C & 0.33528204828(5) & 0.681395(2) &  1.354(10) & 1.358(10) & DNS & (6) \\
B1913+16 & 0.322997448911(4) & 0.6171334(5) & 1.3886(2) & 1.4398(2) & DNS & (7) \\
J0514-4002E & 7.4478966582(72) & 0.70793232(85) & 2.35(19) & 1.53(19) & DNS or NS+BH & (8) \\
J0453+1559 & 4.072468588(4) &  0.11251832(4) & 1.174(4) & 1.559(5) & DNS & (9) \\   
J1918-0642 & 10.9131775801(2) & 0 & 0.219(12) & 1.19(10) & MSP & (10) \\    
J2222-0137 & 2.4457599929(3) & 0 & 1.05(6) & 1.20(14) & MSP & (11) \\    
J1802-2124 & 0.698889243381(5) & 0 & 0.78(4) & 1.24(11) & MSP & (12) \\    
J1713+0747 & 67.82513826930(19) & 0 & 0.286(12) & 1.31(11) & MSP & (13) \\    
J1807-2500B & 9.9566681588(27) & 0.747033198(40) & 1.2064(20) & 1.3655(21) & MSP & (14) \\  
J2234+0611 & 32.001401626(8) & 0.129274035(5) & 0.300(13) & 1.38(1) & MSP & (15) \\   
J0437-4715 & 5.7410459(4) & 0 & 0.224(7) & 1.44(7) & MSP & (16) \\    
J1738+0333 & 0.35479(1) & 0 & 0.181(6) & 1.47(7) & MSP & (17) \\    
J0621+1002 & 8.3186805(1) & 0.00245744(4) & 0.76(18) & 1.53(15) & MSP & (18) \\  
J1909-3744 & 1.533449451246(8) & 0 & 0.214(3) & 1.55(3) & MSP & (19) \\    
J0751+1807 & 0.263144270792(7) & 0 & 0.16(1) & 1.64(15) & MSP & (20) \\    
J1903+0327 & 95.174118753(14) & 0.436678409(3) & 1.029(8) & 1.667(21) & MSP & (21) \\  
J0348+0432 & 0.102424062722(7) & 0 & 0.154(3) & 1.806(37) & MSP & (22) \\   
J1946+3417 & 27.01994783(5) & 0.134495376(10) & 0.2659(11) & 1.828(22) & MSP & (23) \\  
J1012+5307 & 0.604672722901(13) & 0 & 0.16(2) & 1.83(11) & MSP & (24) \\    
J1614-2230 & 8.68661942171(9) & 0 & 0.500(6) & 1.928(17) & MSP & (25) \\    
J2215+5135 & 0.1725021049(8) & 0 & 0.33(2) & 1.98(8) & MSP & (26) \\   
J0740+6620 & 4.76694461933(8) & 0 & 0.251(5) & 2.08(7) & MSP & (27) \\    
B1957+20 & 0.3819666 & 0 & 0.035(2) & 2.40(12) & MSP & (28) \\   
\noalign{\smallskip}\hline
\end{tabular}
\end{center}
\end{table*}

\addtocounter{table}{-1}
\begin{table*}
\caption{\bf Continued.}
\centering
\setlength{\tabcolsep}{0.05cm}
\renewcommand\arraystretch{0.9}
\begin{center}
\begin{tabular}{lcccccc}
\hline\noalign{\smallskip}
Name & $P$ & $e$ & $M_{\rm c}$ & $M_{\rm NS}$ & Type & References \\
 & (day) &  & ($M_{\odot}$) & ($M_{\odot}$) &  \\
\hline\noalign{\smallskip}
J1748-2021B & 20.5500072(6) &  0.5701606(15) & 0.18(29) & 2.74(21) & MSP & (29) \\
4U 1538-522 & 3.728382(11) & 0.18(1) & 16(2) & 1.02(17) & HMXB & (30) \\   
Her X-1 & 1.70016772(1) & 0 & 2.03(37) & 1.07(36) & HMXB & (30) \\   
SMC X-1 & 3.89220909(4) & 0.00089(6) & 18(2) & 1.21(12) & HMXB & (30) \\   
XTE J1855-026 & 6.0724(9) & 0.04(2) & 21(2) & 1.41(24) & HMXB & (30) \\   
LMC X-4 & 1.40839776(26) & 0.006(2) & 18(1) & 1.57(11) & HMXB & (30) \\  
Cen X-3 & 2.087113936(7) & $<$0.0016 & 24(1) & 1.57(16) & HMXB & (30) \\   
SAX J1802.7-2017 & 4.5696(9) & $<$0.2 & 22(2) & 1.57(25) & HMXB & (30) \\  
OAO 1657-415 & 10.44749(55) & 0.1033(6) & 17.5(8) & 1.74(30) & HMXB & (30) \\   
EXO 1722-363 & 9.7403(4) & $<$0.19 & 18(2) & 1.91(45) & HMXB & (30) \\   
Vela X-1 & 8.964368(40) & 0.0898(12) & 26(1) & 2.12(16) & HMXB & (30) \\   
J1123 & 0.27383544(18) & 0 & 0.61(2) & 1.24(3) & Quiescent NS+MS & (31) \\
J1449+6919 & 632.65(21) & 0.2668(10) & 0.91(5) & 1.258(32) & Quiescent NS+MS & (32) \\ 
J0742-4749 & 497.6(4) & 0.168(4) & 0.90(5) & 1.28(4) & Quiescent NS+MS & (32) \\
J0152-2049 & 536.14(18) & 0.6615(10) & 0.782(3) & 1.291(24) & Quiescent NS+MS & (32) \\ 
J06163552 & 0.866605(1) & 0.007(6) & 1.69(13) & 1.30(10) & Quiescent NS+MS & (33) \\
J2057-4742 & 230.15(7) & 0.3095(26) & 1.048(31) & 1.31(4) & Quiescent NS+MS & (32) \\
J1553-6846 & 310.17(11) & 0.5314(21) & 1.04(5) & 1.323(32) & Quiescent NS+MS & (32) \\
J0553-1349 & 189.10(5) & 0.3879(7) & 0.98(6) & 1.33(5) & Quiescent NS+MS & (32) \\
J0003-5604 & 561.83(29) & 0.795(5) & 0.802(3) & 1.34(4) & Quiescent NS+MS & (32) \\
J1733+5808 & 570.94(31) & 0.3093(10) & 1.16(5) & 1.362(30) & Quiescent NS+MS & (32) \\  
J0036-0932 & 719.8(9) & 0.3993(21) & 0.94(4) & 1.362(34) & Quiescent NS+MS & (32) \\
J1739+4502 & 657.4(6) & 0.6777(18) & 0.781(3) & 1.38(4) & Quiescent NS+MS & (32) \\ 
J1150-2203 & 631.81(22) & 0.552(4) & 1.18(6) & 1.39(4) & Quiescent NS+MS & (32) \\
J0217-7541 & 636.1(7) & 0.3228(33) & 0.996(33) & 1.396(33) & Quiescent NS+MS & (32) \\
J2145+2837 & 889.5(7) & 0.5840(35) & 0.95(5) & 1.396(35) & Quiescent NS+MS & (32) \\  
J0230+5950 & 1029(5) & 0.753(11) & 1.114(30) & 1.401(34) & Quiescent NS+MS & (32) \\ 
J1255 & 2.760541(13) & 0 & 1.13(5) & 1.41(20) & Quiescent NS+MS & (34) \\
\noalign{\smallskip}\hline
\end{tabular}
\end{center}
\end{table*}

\addtocounter{table}{-1}
\begin{table*}
\caption{\bf Continued.}
\centering
\setlength{\tabcolsep}{0.2cm}
\renewcommand\arraystretch{0.9}
\begin{center}
\begin{tabular}{lcccccc}
\hline\noalign{\smallskip}
Name & $P$ & $e$ & $M_{\rm c}$ & $M_{\rm NS}$ & Type & References \\
 & (day) &  & ($M_{\odot}$) & ($M_{\odot}$) &  \\
\hline\noalign{\smallskip}
J2244-2236 & 938.3(5) & 0.5666(11) & 1.002(30) & 1.443(23) & Quiescent NS+MS & (32) \\
J2102+3703 & 481.04(26) & 0.448(9) & 1.03(3) & 1.473(34) & Quiescent NS+MS & (32) \\
J0634+6256 & 1046.0$\pm$2.1 & 0.564(11) & 1.18(6) & 1.48(9) & Quiescent NS+MS & (32) \\
J2354 &  0.47992(1) & 0.002(2) & 0.73(6) & 1.50(10) & Quiescent NS+MS & (35) \\  
J1048+6547 & 827(5) & 0.357(9) & 0.99(5) & 1.52(7)  & Quiescent NS+MS & (32) \\
J0824+5254 & 1026.7$\pm$3.3 & 0.686(12) & 1.102(30) & 1.604(34) & Quiescent NS+MS & (32) \\
J0639-3655 & 654.6(6) & 0.721(13) & 1.32(6) & 1.70(7) & Quiescent NS+MS & (32) \\
{\it Gaia} NS1 & 730.8(12) & 0.123(2) &  0.79(1) & 1.90(4) & Quiescent NS+MS & (36) \\  
\noalign{\smallskip}\hline
\end{tabular}
\end{center}
References: (1) \citep{2008PhRvD..77l4017B} and \citep{2011MNRAS.412..580A}, (2) \citep{2006ApJ...640..428L} and \citep{2015ApJ...798..118V}, (3) \citep{1991Natur.350..688W} and \citep{2014ApJ...787...82F}, (4) \citep{2003Natur.426..531B} and \citep{2006Sci...314...97K}, (5) \citep{2005ApJ...618L.119F} and \citep{2014MNRAS.443.2183F}, (6) \citep{1989IAUC.4772....1A} and \citep{2006ApJ...644L.113J}, (7) \citep{1975ApJ...195L..51H} and \citep{2010ApJ...722.1030W}, (8) \citep{2024Sci...383..275B}, (9) \citep{2013ApJ...775...51D} and \citep{2015ApJ...812..143M}, (10) \citep{2016PhDT.......158F}, (11) \citep{2014ApJ...789..119K}, (12) \citep{2010ApJ...711..764F}, (13) \citep{2015ApJ...809...41Z}, (14) \citep{2012ApJ...745..109L}, (15) \citep{2019ApJ...870...74S}, (16) \citep{2016MNRAS.455.1751R}, (17) \citep{2012MNRAS.423.3316A}, (18) \citep{2012PhDT.........4K}, (19) \citep{2016PhDT.......158F} and \citep{2016MNRAS.458.3341D}, (20) \citep{2016MNRAS.458.3341D}, (21) \citep{2011MNRAS.412.2763F}, (22) \citep{2025ApJ...983L..20S}, (23) \citep{2017MNRAS.465.1711B}, (24) \citep{2016arXiv160501665A}, (25) \citep{2016ApJ...832..167F}, (26) \citep{2018ApJ...859...54L} and \citep{2024ApJ...974..315S}, (27) \citep{2021ApJ...915L..12F}, \citep{2020NatAs...4...72C} and \citep{2019MNRAS.485.3715B}, (28) \citep{2011ApJ...728...95V}, (29) \citep{2008ApJ...675..670F}, (30) \citep{2003ApJ...596L..63A}, \citep{2015A&A...577A.130F} and \citep{2011ApJ...730...25R}, (31) \citep{2022NatAs...6.1203Y}, (32) \citep{2024OJAp....7E..58E}, (33) \citep{2022ApJ...940..165Y}, (34) \citep{2022MNRAS.517.4005M}, (35) \citep{2023SCPMA..6629512Z}, (36) \citep{2024OJAp....7E..27E}.
\end{table*}

\clearpage

\end{document}